\documentclass[aip,notitlepage, floatfix, jcp, reprint]{revtex4-1}%
\usepackage{amssymb}
\usepackage{amsfonts}
\usepackage{amsmath}
\usepackage{graphicx}
\usepackage{enumerate}

\usepackage[usenames,dvipsnames]{color}
\usepackage{ulem}%
\providecommand{\U}[1]{\protect\rule{.1in}{.1in}}

\begin{document}

\title{Extending classical nucleation theory to confined systems}
\author{Miguel A. Dur\'{a}n-Olivencia}
\affiliation{\label{IACT}Laboratorio de Estudios Cristalogr\'{a}ficos. Instituto Andaluz de
Ciencias de la Tierra, CSIC-UGR, Code Postal 18100, Avenida de las Palmeras, 4 Granada, Spain}
\email{maduran@lec.csic.es}
\author{James F. Lutsko}
\affiliation{\label{CENOLI}Center for Nonlinear Phenomena and Complex Systems, Code Postal 231,
Universit\'{e} Libre de Bruxelles, Blvd. du Triomphe, 1050 Brussels, Belgium}
\email{jlutsko@ulb.ac.be}
\homepage{http://www.lutsko.com}

\begin{abstract}
Classical nucleation theory has been recently reformulated
 based on fluctuating hydrodynamics
 [J.F. Lutsko and M.A. Dur\'{a}n-Olivencia, J. Chem. Phys.  {\bf 138},
244908(2013)]. The present work extends this effort to the case of nucleation
in confined systems such as small pores and vesicles. The finite available 
mass imposes a maximal supercritical cluster size and prohibits nucleation 
altogether if the system is too small.  We quantity the effect of system size on 
the nuceation rate. We also discuss the effect of relaxing the capillary-model 
assumption of zero interfacial width resulting in significant changes in the nucleation 
barrier and nucleation rate.   
\end{abstract}
\date{\today }
\maketitle

\section{Introduction\label{sec:introduction}}

Nucleation is a ubiquitous process in nature which has been the subject of
extensive research throughout the last century. Nowadays the most well-known
understanding of the process relies on Gibbs'
work\cite{article:gibbs-seminal-1878-a,article:gibbs-seminal-1878-b,
book:gibbs-1931} concerning the characterization of phase transformations. Mainly
focused on transitions near the equilibrium, Gibbs deduced a simple 
expression for the work required to form a spherical embryo (so-called
cluster) of the new phase within the old one, $W(r)$ with $r$ being the
cluster radius. While these efforts set the thermodynamic ground for
understanding nucleation phenomena,
\citeauthor{article:volmer-weber-1926}\cite{article:volmer-weber-1926,
book:volmer-1939}
were the pioneers to reveal the importance of the kinetics of nucleation. They
proposed a rudimentary model to account for the chief characteristics of such
phenomenon. A short time later, a more atomistic picture was proposed by
\citet{article:farkas-1927} who developed the idea of Szilard and that was
further developed by \citet{article:becker-doring-1935} resulting in the
equation which now bears their names. Finally,
\citeauthor{book:frenkel-1946}\cite{article:frenkel-1939,book:frenkel-1946} and
\citet{article:zeldovich-1943} reached a similar result which also allows to
describe
non-steady-state kinetics. \citet{article:turnbull-fisher-1949} generalized this 
formalism in order to describe solid nucleation from a liquid phase,
an approach that was readily extended to include nucleation in solids. The
nucleation rate expressions derived from all these developments have an
Arrhenius-like structure\cite{book:kashchiev-2000,book:kelton-2010} but they
differ in the exact expression for the pre-exponential factor. The combination
of these ideas comprise a remarkably robust theory which is commonly
called {Classical Nucleation Theory} (CNT).

Besides being a versatile tool, CNT is intuitively appealing and
clearly summarizes the basic rules underlying phase transformations.
However, while CNT has shown an extraordinary ability to predict the functional
dependence of the nucleation rate on the thermodynamic variables involved, it
has exhibited a severe disability when it comes to quantitatively explain
experimental data.
\cite{article:viisanen-strey-reiss-1993,article:viisanen-strey-1994,article:hruby-viisanen-strey-1996}
This flaw has been usually blamed either on a poorly refined expression of the
work of cluster formation, or on the heuristic modelling of cluster formation
based on macroscopic growth laws, or on the simplicity of the cluster properties
assumed by the capillary approach. There has been several attempts to extend and
refine CNT, e.g. generalizing  the kinetic
model\cite{article:shizgal-barret-1989,article:kashchiev-1969b} to consider
wider cluster transitions than that initially assumed by the pioneers of
nucleation,\cite{article:becker-doring-1935,
article:kaichew-stranski-1934,article:zeldovich-1943,
article:frenkel-1939,book:frenkel-1946,
article:tunitskii-1941} providing more accurate expressions of the free energy
barrier by using classical Density Functional
Theory\cite{inbook:kelton-1991,article:lutsko-2011-b} (DFT), refining the
capillary model,\cite{article:prestipino-2012} or by selecting
a different order parameter instead of the cluster size to characterize the
nucleation pathway.\cite{article:lechner-2011} Recently, a new approach to nucleation has been formulated\cite{article:lutsko-2011-c,article:lutsko-2012-dtn,
article:lutsko-2012-a1} based on fluctuating hydrodynamics.\cite{book:landau-lifshitz-fluidMechanics-1959} We will refer to this as Mesoscopic Nucleation Theory (MeNT).
This new framework provides a self-consistent
justification and extension of more heuristic equilibrium approaches based
solely on the free energy. The MeNT provides a general stochastic differential
equation (SDE) for the evolution of an arbitrary number of order parameters
characterizing the number density field. When the simplest case is considered,
that is a single order parameter, a straight-forward connection with CNT is
found.\cite{article:lutsko-duran-2013} Such a reformulation of CNT, hereafter
called dynamical CNT (dCNT), sheds light on the weaknesses of the classical
derivation and can be used to construct a more realistic theory in which clusters
have finite interfacial width.

The present work aims to continue this development so as to extend dCNT to the case of  confined
systems. In the last few years there has been a veritable explosion of interest
in nucleation due to the development of new techniques, such as
microfluidics, that
bring us the opportunity to probe the very small and the very fast.  
Besides, nucleation in confined environments is important for biological processes such as
bone formation,\cite{article:meldrum-2013bone,article:delgado-lopez-2013apatite} \emph{in vivo} protein 
crystallization,\cite{article:bechtel-1976parasporal,article:koopmann-2012invivo} or 
cavitation in lipid bilayers,\cite{article:renn-2013cavitation} to name but a few.
However,
CNT is based on assumptions that are violated for small systems. For example, 
when the nucleation of a dense droplet from a weak solution
is considered it is assumed that clusters do not consume
enough material during  nucleation so as to have a noticeable effect on the properties 
of the mother phase, but
this can only be true for large systems. The main goal of this
work is therefore to extend dCNT to take into consideration the
conservation of mass required for a finite volume with the aim of further
developing the classical theory. Following a similar procedure
as that presented in a previous work,\cite{article:lutsko-duran-2013} a
nucleation rate equation is readily obtained. It turns out that the confinement
has a strong effect on the energy barrier and, thus, on the nucleation
rate.
On the one hand, in contrast to infinite systems, the cluster of new phase 
can only grow to a certain maximal size so that a complete phase transition is not possible. 
Nevertheless, for sufficiently large and supersaturated systems, this maximal 
cluster is indeed the stable, equilibrium state.  In contrast, if the system is too small, 
the maximal cluster size is less than the critical radius and no transition takes place.
In other words, nucleation is found to be inhibited as a consequence of the size of
the container where the experiment is being carried out.
On the other hand, nucleation rate is affected for a certain range of
volumes when we compare it with the CNT prediction calculated for 
infinite systems.\cite{article:lutsko-duran-2013} Indeed, such a ratio
shows a maximum for system sizes close to that which inhibits nucleation.
Moreover, considerable corrections arise when
a more realistic model for clusters is taken into consideration.

In section \ref{sec:theory} the order-parameter dynamics derived from fluctuating hydrodynamics is
modified so that the finite volume limit is taken into account. It is shown that the confinement
does not affect the structure of the stochastic differential equation (SDE) derived in
Ref. \onlinecite{article:lutsko-2012-dtn}. The use of this SDE with
a modified version of the capillary model that accounts for the finite mass in the system under study
is presented in section \ref{sec:capillaryModel}. In that Section, we give expressions
for the attachment rate, the stationary cluster-size distribution, the nucleation rate and the growth rate
of super-critical clusters. Section \ref{sec:extendedModels} focuses on the improvement of those
results by means of considering clusters with a finite interfacial width. Three
models are proposed: in the first the inner density and the interfacial width are the same as in the case
of infinite systems, in the second the inner density is chosen so as to minimize the free energy of the stable
cluster, and lastly, in the third model, both the interior density and the interfacial width
are determined so as to minimize the free energy of the stable cluster.
While the first two models yield similar results between them and to the capillary approach, 
the last one gives rise to large deviations from the other models.
These comparisons are presented in section \ref{sec:resultsAndComparisons}.
Finally, our results are summarized in section \ref{sec:conclusions}.

\section{Theory\label{sec:theory}}

The approach we follow in this work, based on Ref. \onlinecite{article:lutsko-2012-dtn}, requires that a spherical cluster be characterized by its density as a function of distance from its center, $\rho(r;\mathbf{x}(t))$ where $\mathbf{x}$ represents a set of one or more parameters describing the cluster: e.g. its radius, interior density, etc. As indicated, these parameters can change in time according to the following SDE,
\cite{article:lutsko-2011-c,article:lutsko-2012-dtn,
article:lutsko-2012-a1} which will be the basis for this study,
\begin{align}
 \frac{dm(r;\mathbf{x}(t))}{dt}=&\,D4\pi r^2\rho(r;\mathbf{x}(t))\left.\frac{\partial}{\partial r}
 \frac{\delta\beta F[\rho]}{\delta\rho(\mathbf{r})}\right|_{\rho(r;\mathbf{x}(t))} \nonumber\\
 &-\sqrt{D8\pi r^2\rho(r;\mathbf{x}(t))}\,\xi(r;t)\label{eq:cumulative-mass-SDE},
\end{align}
where $m(r;\mathbf{x}(t))$ stands for the mass inside a spherical shell,
 \begin{equation}
  m(r;\mathbf{x}(t))=4\pi\int_{0}^{r} \rho(r';\mathbf{x}(t))\,r'^2dr'
  \label{eq:definition-cumulative-mass}
 \end{equation}
and, with $D$ being the diffusion constant, $F[\rho]$ being the Helmholtz free energy, $\beta=1/k_BT$
 where $k_B$ is the Boltzmann constant and $T$ is the absolute temperature, and where $\xi(r;t)$ 
 is a fluctuating force
that fulfils
\begin{equation}
 \langle\xi(r;t)\xi(r';t')\rangle=\delta(r-r')\delta(t-t').
\end{equation}
Note that square brackets in equation (\ref{eq:cumulative-mass-SDE}) have been used to indicate a  functional dependence.
Finally,  it has been shown that equation (\ref{eq:cumulative-mass-SDE}) is It\^{o}-Stratonovich equivalent
(see appendix A of Ref. \onlinecite{article:lutsko-2012-dtn}), for which reason so that either interpretation may be used.

The next step consists of deriving the dynamics of the parameter vector, $\mathbf{x}(t)$, in confined volumes which will open the door to reduced descriptions, specifically to single order-parameter description.

\subsection{Order-parameter dynamics in confined systems
\label{subsec:orderParameterDynamicsInConfinedSystems}}

The use of a finite number of scalar parameters (so-called order parameters) to describe the 
 density can be a crude simplification but it is also a very useful method to get an approximate representation
of the whole problem. Such a reduced description of the real density profile is commonly used in the classical
picture, where it is customary to hypothesize that density fluctuations are well characterized by a single order parameter,
namely the size of the cluster. While in CNT the order-parameter dynamics is formulated based on heuristic reasoning, MeNT allows
us to derive the dynamical equations from a formal point of view, including  the case of more than one order parameter. Here, we briefly review the 
arguments leading to the equations for the order parameter in order to note the effect of imposing a finite volume.

From equation (\ref{eq:cumulative-mass-SDE}) the time-evolution equation governing the order-parameter dynamics is given by,
\begin{align}
 \frac{\partial m(r;\mathbf{x}(t))}{\partial x_i}\frac{d x_i}{dt}=&\,D4\pi r^2\rho(r;\mathbf{x}(t))\left.\frac{\partial}{\partial r}
 \frac{\delta\beta F[\rho]}{\delta\rho(\mathbf{r})}\right|_{\rho(r;\mathbf{x}(t))} \nonumber\\
 &-\sqrt{D8\pi r^2\rho(r;\mathbf{x}(t))}\,\xi(r;t)\label{eq:xi-cumulative-mass-SDE}.
\end{align}
Now, let us assume that the container is a sphere of radius $R_T$. The line of reasoning 
presented in section III.B of Ref. (\onlinecite{article:lutsko-2012-dtn}) remains valid, although we have 
to take care of imposing the right integration limits in order to consider the confinement.
 Thus, the latter equation can be transformed into equation
(\ref{eq:Wj-xi-SDE}) multiplying by a function $W_j(r;\mathbf{x}(t))$ and integrating up to $R_T$,
\begin{align}
 g_{ij}(\mathbf{x})\frac{d x_i}{dt}=&\,D\int_0^{R_T} W_j(r;\mathbf{x}(t))\rho(r;\mathbf{x}(t))\times\nonumber\\
 &\times\left(\frac{\partial}{\partial r}
 \frac{\delta\beta F[\rho]}{\delta\rho(\mathbf{r})}\right)_{\rho(r;\mathbf{x}(t))} 4\pi r^2dr\nonumber\\
 &-\int_0^{R_T} W_j(r;\mathbf{x}(t)) \sqrt{D8\pi r^2\rho(r;\mathbf{x}(t))}\,\xi(r;t)dr\label{eq:Wj-xi-SDE},
\end{align}
with
\begin{equation}
 g_{ij}(\mathbf{x}(t))=\int_0^{R_T}W_j(r;\mathbf{x}(t))\frac{\partial m(r;\mathbf{x}(t))}{\partial x_i}dr
 \label{eq:definition-general-gij}
\end{equation}
It was shown that if the diffusion matrix $\mathcal{D}_{ij}(\mathbf{x})$ associated to equation (\ref{eq:Wj-xi-SDE}) and
the matrix $g_{ij}(\mathbf{x})$ are assumed proportional, the function $W_i$ must be, modulo a multiplicative constant,
\begin{equation}
 W_i(r;\mathbf{x}(t))=\frac{1}{4\pi r^2\rho(r;\mathbf{x}(t))}\frac{\partial m(r;\mathbf{x}(t))}{\partial x_i}
 \label{eq:definition-Wi}
\end{equation}
so that $\mathcal{D}_{ij}(\mathbf{x})=2Dg_{ij}(\mathbf{x})$ and, eventually,
\begin{equation}
 g_{ij}(\mathbf{x})=\int_0^{R_T}\frac{1}{4\pi r^2\rho(r;\mathbf{x}(t))}\frac{\partial m(r;\mathbf{x}(t))}{\partial x_i}\frac{\partial m(r;\mathbf{x}(t))}{\partial x_j}dr,
\end{equation}
which is also called ``the metric''\cite{article:lutsko-2011-c,article:lutsko-2012-dtn,article:lutsko-duran-2013}. The inverse of this matrix will be seen below to be interpretable as the matrix of state-dependent kinetic coefficients. By using the definition of $W_i(\mathbf{x})$  (Eq. \ref{eq:definition-Wi}), the equation for the driving force (\ref{eq:Wj-xi-SDE}) becomes,
\begin{align}
 \int_0^{R_T}& W_j(r;\mathbf{x}(t))\rho(r;\mathbf{x}(t))\left.\frac{\partial}{\partial r}
 \frac{\delta\beta F[\rho]}{\delta\rho(\mathbf{r})}\right|_{\rho(r)} 4\pi r^2dr\nonumber\\
 =&\left[\frac{\partial m(r;\mathbf{x}(t))}{\partial x_j}\left.\frac{\delta\beta F[\rho]}{\delta\rho(\mathbf{r})}\right|_{\rho(r;\mathbf{x}(t))}\right]_0^{R_T}\nonumber\\
 &-\int_{r<R_T}\frac{\partial \rho(r;\mathbf{x}(t))}{\partial x_j}\left.\frac{\delta\beta F[\rho]}{\delta\rho(\mathbf{r})}\right|_{\rho(r;\mathbf{x}(t))}d\mathbf{r}.\label{eq:thermo-driving-force-proof-1}
\end{align}
The first term gives a zero contribution at $r=0$, and at $r=R_T$ the contribution will be,
\begin{equation}
 \frac{\partial m(r;\mathbf{x}(t))}{\partial x_j}\left.\frac{\delta\beta F[\rho]}{\delta\rho(\mathbf{r})}\right|_{\rho(R_T;\mathbf{x}(t))}=\frac{\partial N}{\partial x_j}\,\mu(\rho(R_T;\mathbf{x}(t))),
\end{equation}
which vanishes in closed systems for which the total number of particles, $N=m(R_T;\mathbf{x})$, is constant regardless the values of the order parameters. The second term in equation (\ref{eq:thermo-driving-force-proof-1}) can be simplified by using the functional chain rule,
\begin{equation}
 \int_{r<R_T}\frac{\partial \rho(r;\mathbf{x}(t))}{\partial x_j}\left.\frac{\delta\beta F[\rho]}{\delta\rho(\mathbf{r})}\right|_{\rho(r;\mathbf{x}(t))}d\mathbf{r} = \frac{\partial\beta F(\mathbf{x})}{\partial x_j},
\end{equation}
where $F(\mathbf{x})$ has been used as the equivalent of $F[\rho]$. The latter equations allows to rewrite the driving-force term of
the SDE (\ref{eq:Wj-xi-SDE}) in a simpler manner that involves only partial derivatives. The noise term is similarly simplified following Ref.(\onlinecite{article:lutsko-2012-dtn}) to get 
\begin{equation}
\frac{dx_i}{dt}=-Dg_{ij}^{-1}(\mathbf{x})\frac{\partial\beta F(\mathbf{x})}{\partial x_i}+2DA_i(\mathbf{x})-\sqrt{2D}q_{ji}^{-1}(\mathbf{x})\xi(t)
\label{eq:order-parameter-sde-simplified}
\end{equation}
with $q_{il}(\mathbf{x})q_{jl}(\mathbf{x})=g_{ij}(\mathbf{x})$ and
\begin{align}
A_{i}\left( \mathbf{x}\right) =&q_{ik}^{-1}\left( \mathbf{x}\right)
\frac{\partial q_{jk}^{-1}\left( \mathbf{x}\right) }{\partial x_{j}}-
\frac{1}{2}g_{il}^{-1}\left( \mathbf{x}\right) \frac{\partial g_{jm}^{-1}
\left( \mathbf{x}\right) }{\partial x_{l}}g_{mj}\left( \mathbf{x}\right)\nonumber\\
&+\frac{1}{2}\left( g_{il}^{-1}\left( \mathbf{x}\right)
g_{jm}^{-1}\left(\mathbf{x}\right) -g_{ij}^{-1}\left( \mathbf{x}\right)
g_{lm}^{-1}\left(\mathbf{x}\right) \right)\nonumber\\
&\times\int_{0}^{R_T}\frac{1}{4\pi r^{2}\rho^{2}\left( r;\mathbf{x}\right) }
\left(\frac{\partial \rho \left( r;\mathbf{x}\right) }{\partial x_{l}}\right.\nonumber\\
&\times\left.\frac{\partial m
\left( r;\mathbf{x}\right) }{\partial x_{j}}\frac{\partial m
\left(r;\mathbf{x}\right) }{\partial x_{m}}\right)dr \label{eq:Ai-definition}.
\end{align}
This has exactly the same in structure as the counterpart for open systems except that the free energy that occurs here is the Helmholtz free energy while for open systems it is, naturally enough, the grand potential. Hence, the confinement does not alter the structure of the dynamics equations, as expected,
but it will play an important role when it comes to derive the exact expressions of
the cluster density profile, the free energy and the cumulative mass.

This framework will be applied to make contact with the classical picture
but considering a finite mass and volume. The following sections are intended to modify the
capillary and extended models discussed by \citet{article:lutsko-duran-2013} by enforcing
the mass conservation law,
\begin{equation}
 N=4\pi \int_0^{R_T} \rho(r;\mathbf{x}(t))\ r^2dr.
 \label{eq:mass-conservation-law}
\end{equation}
where $N$ represents the total number of particles, also referred as the ``total mass'', which is strictly constant for a closed system.
To this end we will particularize the general order-parameter dynamics  to a  single order-parameter
description, i.e. a one-dimensional parametrization will be considered.
In contrast to CNT, the chosen parameter may be indifferently the cluster size in number of molecules or in radius,
or even an abstract variable to simplify the resulting SDE. Hereinafter we will also specialize to the case that the new phase is more dense than the old phase (e.g. nucleation of liquid from gas) although the opposite possibility (e.g. nucleation of gas from liquid) is very similar.

\subsubsection{One-dimensional parametrization}

For the simplest case of a single order parameter, 
\begin{equation}
 \rho(r;t)\rightarrow \rho(r;X(t))\label{eq:general-1-dimensional-rho}.
\end{equation}
it was shown\cite{article:lutsko-2011-c,article:lutsko-2012-dtn,article:lutsko-2012-a1}
that equation (\ref{eq:order-parameter-sde-simplified}) becomes,
\begin{align}
 \frac{dX}{dt}=&-Dg^{-1}(X)\frac{\partial \beta F(X)}{\partial X}-D\frac{1}{2}g^{-2}(X)\frac{\partial g(X)}{\partial X}\nonumber\\
 &+\sqrt{2D\,g^{-1}(X)}\xi(t),
 \label{eq:dynamics-dCNT}
\end{align}
which constitutes the starting point of the dCNT. The metric in this reduced description
is a 1-dimensional function of $X$ whose definition according to equation
(\ref{eq:definition-general-gij}) becomes,
\begin{equation}
 g(X)=\int_0^{R_T} \frac{1}{4\pi r^2\rho(r;X)}\left(\frac{\partial m(r;X)}{\partial X}\right)^2 dr.
 \label{eq:definition-1dgeneral-gij}
\end{equation}
As for the cumulative mass, the definition (\ref{eq:definition-cumulative-mass}) remains
unchanged but now $\mathbf{x}(t)=X(t)$.
That said, equation (\ref{eq:order-parameter-sde-simplified}) is easily transformed into a
Fokker-Planck equation (FPE) determining the time evolution of the probability density
function (PDF) of the random variable $X$,\cite{book:risken-1996a,book:gardiner-2004,
article:lutsko-2012-dtn,article:lutsko-duran-2013}
\begin{align}
 \frac{\partial P(X,t)}{\partial t}=&-\frac{\partial \mathfrak{J}(X,t)}{\partial X},
 \label{eq:fpe-general-x}
\end{align}
with
\begin{align}
 \mathfrak{J}(X,t)=&\, -D\left(\nonumber g^{-1}(X)\frac{\partial \beta F(X)}{\partial X}\right.\nonumber\\
 &+\left.g^{-1/2}(X)\frac{\partial }{\partial X}g^{-1/2}(X)\right)P(X,t)\label{eq:fpe-general-v1-x}\\
 =&-D\left(\nonumber g^{-1}(X)\frac{\partial  \left(\beta F(X)-\ln g^{1/2}(X)\right)}{\partial X}\right.\nonumber\\
 &+\left.g^{-1}(X)\frac{\partial }{\partial X}\right)P(X,t)\label{eq:fpe-general-v2-x}
\end{align}
being the probability flux, which has been written in two ways to show the similarity
with the Zeldovich-Frenkel
equation\cite{article:zeldovich-1943,article:frenkel-1939,book:frenkel-1946} of CNT.
Indeed, the FPE determined by equations (\ref{eq:fpe-general-x}, \ref{eq:fpe-general-v2-x})
is formally equivalent to the Zeldovich-Frenkel
equation when $X$ is the number of molecules inside a cluster,
with $Dg^{-1}(X)$ playing the role of the monomer-attachment rate and the free energy
shifted by a logarithmic term in $g(X)$. It has been shown\cite{article:lutsko-duran-2013}
that the logarithmic term ensures
the general covariance of the dCNT. This means that when different equivalent choices of the parmater $X(t)$ are possible (e.g. the mass or radius of the cluster), the stochastic dynamics will be independent of which parameter is used - a nontrival property that does not occur naturally in the context of CNT. 

While the general solution of equation (\ref{eq:fpe-general-x}) is a difficult problem,
a simple case admitting a solution is that of a stationary system with const flux, $\mathfrak{J}_s$,
so that,
\begin{align}
 \mathfrak{J_s}=&-D\left(\nonumber g^{-1}(X)\frac{\partial \beta F(X)}{\partial X}\right.\nonumber\\
 &+\left.g^{-1/2}(X)\frac{\partial }{\partial X}g^{-1/2}(X)\right)P(X),
\end{align}
from which we readily obtain,
\begin{align}
P_{s}(X)=&A\,g^{1/2}(X)e^{-\beta F(X)}\nonumber\\
&-\frac{\mathfrak{J}_s}{D}g^{1/2}(X)e^{-\beta F(X)}\int^X g^{1/2}(Y)e^{\beta F(Y)} dY,
\label{eq:steady-state-pdf}
\end{align}
the steady-state solution, where $A$ is a normalization constant and
which is manifestly invariant under transformation
of variables.\cite{article:lutsko-duran-2013} 
If we consider that such a stationary non-zero flux is ensured by removing 
clusters once they reach a given size $X_+$, the steady-state distribution 
must satisfy $P_s(X_+)=0$. When this
condition is imposed, equation (\ref{eq:steady-state-pdf}) becomes,
\begin{align}
P_{s}(X)=&\ \frac{\mathfrak{J}_s}{D}g^{1/2}(X)e^{-\beta F(X)}\int_X^{X_+} g^{1/2}(X')e^{\beta F(X')} dX'.
\label{eq:steady-state-final-pdf}
\end{align}

For an undersaturated solution, equilibrium, of course, can be identified
with a particular value of the stationary flux, namely $\mathfrak{J}_s=0$. Thus,
when the system is in a equilibrium state (i.e., under-saturated) the PDF will
be,
\begin{align}
 P_{eq}(X)&= A g^{1/2}(X)\exp\left(-\beta F(X)\right)\nonumber\\
 &=A\exp\left\{-\beta \left(F(X)-\frac{1}{2}k_BT\ln g(X)\right)\right\}.
 \label{eq:PDF-equilibrium}
\end{align}

\subsubsection{Canonical form: the natural order parameter}

Thus far, our concern was to use  the mathematical tools of the theory of
stochastic processes in order to make contact with CNT, what led us to
derive a formally equivalent to the Zeldovich-Frenkel
equation. However, any single-variable SDE
with multiplicative noise (as the current case) can be always transformed
into a simpler one with additive noise via the transformation of variable,
\cite{book:risken-1996a,book:gardiner-2004,article:lutsko-duran-2013}
\begin{equation}
 dY = \sqrt{g(X)}\,dX.
\label{eq:canonical-transformation}
\end{equation}
with an arbitrary boundary condition that for the sake of simplicity will be taken to
be $Y(0)=0$. Such a ``canonical variable'' is the most natural order parameter
to be chosen in the case of a one-dimensional parametrization of $\rho(r;t)$,
since equation (\ref{eq:dynamics-dCNT}) is thereby simplified,
\begin{equation}
 \frac{dY}{dt}= -D\frac{\partial \beta \widetilde{F}(Y)}{\partial Y}+\sqrt{2D}\,\xi(t),
 \label{eq:canonical-sde}
\end{equation}
where $\widetilde{F}(Y)=F(X(Y))$. As can be observed, such an equation is It\^{o}-Stratonovich
equivalent. 
The same goes for the FPE (\ref{eq:fpe-general-x}) which becomes,\cite{article:lutsko-duran-2013}
\begin{equation}
 \frac{\partial \widetilde{P}(Y,t)}{\partial t}=D\frac{\partial }{\partial Y}
 \left(\frac{\partial\beta \widetilde{F}(Y)}{\partial Y}+\frac{\partial}{\partial Y}\right)\widetilde{P}(Y,t)
 \label{eq:canonical-fpe}
\end{equation}
with $\widetilde{P}(Y,t)dY=P(X,t)dX$. These equations will be very useful when it comes
to get the nucleation rate since they notably simplify the calculations
involved in the derivation.

\subsubsection{Nucleation rate and mean first-passage time
\label{subsub:nucleationRateAndMFPT}}

In the previous study for infinite systems the nucleation rate  was
derived from classical arguments yielding an expression  that essentially
corroborated the well-known relationship between the nucleation rate and the 
mean first-passage time (MFPT),\cite{book:barrat-hansen-2003} hereafter denoted as $\tau$ 
and accompanied by a subscript to specify the corresponding approach,
\begin{align}
 J_{\text{CNT}} \equiv&\,\frac{\rho_{av}}{2\tau_{\text{CNT}}}=\frac{D\rho_{av}}{\int_{X_1}^{X_{+}}g_\infty(X')e^{\beta\Delta\Omega(X')}dX'}\notag\\
 \sim& \, \rho_{av}\, Dg_\infty^{-1}(\Delta N_*)\,\sqrt{\frac{1}{2\pi}\left|\beta\Delta \Omega''_*\right|}
 \exp\left(-\beta\Delta\Omega_*\right),
\label{eq:tau-cnt-approx}
\end{align}
where the infinite subscript has been used to remember that the metric used here is that
for an infinite system, $\Omega=F-\mu N$ is the grand canonical potential, 
$X_1$ is the value of the order parameter $X$ for which the number of molecules
inside the cluster, $\Delta N$, is set to be 1, $X_+$ can be any value beyond the critical size
to enforce the stationary flux, and where
\begin{align}
 \beta\Delta\Omega_*'\equiv& \beta\Delta\Omega ' (X_*)=0,\nonumber\\
 \beta\Delta\Omega_*''\equiv& \beta\Delta\Omega'' (X_*).
\end{align}
Indeed, the MFPT can be directly identified
as the time required for the phase transition to start, since one super-critical cluster
in the whole system is enough to trigger the transition. Adapting the same argument
as led to  (\ref{eq:tau-cnt-approx}) one can derive the escape rate for confined systems, 
\begin{align}
 j_{nc}&\equiv\frac{1}{\tau_{nc}}= \frac{2D}{\int_{X_1}^{X_{+}}g(X')e^{\beta\Delta F(X')}dX'}\nonumber\\
 &\sim 2D\, g^{-1}(\Delta N_*)\,\sqrt{\frac{1}{2\pi}\left|\beta\Delta F''_*\right|}
 \exp\left(-\beta\Delta F_*\right)
 \label{eq:escape-rate-nc}
\end{align}
Note that in the following we will not distinguish between the escape rate and the
nucleation rate, as they are essentially the same. This will be ulteriorly compared to
the classical estimation (from Eq. \ref{eq:tau-cnt-approx}). Such a ratio will give us
a first idea of the effect of the mass conservation.

Given that we are restricting attention to the evolution of a 
single cluster which is not perturbed by any other clusters within the 
system, it seems natural to focus on the escape rates. In our particular case the MFPT is given by,
\begin{align}
 \tau=\,\frac{1}{2D}&\int_{0}^{X_+}dx\,
		    P_0(x)\int_{x}^{X_{+}}dx'\,g^{1/2}(x')e^{\beta {F}(x')}\times\notag\\
                    &\times\int_0^{x'} dx''\,g^{1/2}(x'')e^{-\beta  F(x'')}.
 \label{eq:MFPT-definition}
\end{align}
Considering the initial PDF, $P_0(X)=\delta(x)$, the latter equation becomes,
\begin{equation}
  \tau = \frac{1}{2D}\int_0^{X_+}dx\,g^{1/2}(x)e^{-\beta{F}(x)}
    \int_{x}^{X_+}dx'\,g^{{1/2}}(x')e^{\beta{F}(x')}
    \label{eq:escape-rate-definition}.
\end{equation}
It is not generally possible to evaluate this expression analytically, however  we can make a good 
approximation of its value with the aid of the canonical variable and assuming 
the free energy admits the expansion,
$
\beta\widetilde{ F}(Y)=\beta\widetilde{F}(Y(0))+\widetilde{F}_0\,Y^\alpha+\dots\
$
 with some $\alpha >0$, so that  it can be approximated as
\begin{align}
\beta\Delta\widetilde{F}(Y)\sim&\ \widetilde{F}_0\,Y^\alpha
\end{align}
for small values of $Y$.  Hence, by using the same method explained in appendix
A of Ref. \onlinecite{article:lutsko-duran-2013}, the escape rate becomes,
\begin{align}
 j=&\ \frac{2D}{\int_0^{X_+}dx\,g^{1/2}(x)e^{-\beta{F}(x)}
    \int_{x}^{X_+}dx'\,g^{{1/2}}(x')e^{\beta{F}(x')}}\notag\\
 \sim&\ 2D \frac{\alpha\,\widetilde{F}_0^{1/\alpha}\left(2\pi|\beta\widetilde{F}''(Y_*)|^{-1}\right)^{-1/2}}{\left(\Gamma\left(\frac{1}{\alpha}\right)-\Gamma_i\left(\frac{1}{\alpha},\widetilde{F}_0Y^{\alpha}(X_+)\right)\right)}{e}^{-\beta\Delta F_*}\nonumber\\
 \sim&\ 2D\,\frac{\alpha\,\widetilde{F}_0^{1/\alpha}}{\Gamma\left(\frac{1}{\alpha}\right)}\,\sqrt{\frac{1}{2\pi}|\beta F''(X_*)|g^{-1}(X_*)}\,e^{-\beta\Delta F_*}
 \label{eq:escape-rate-general}
\end{align}
Note that the tilde has been used to highlight that the expression of the free energy is 
written in terms of the canonical variable. Indeed, this equation can also be deduced
from the dCNT derivation by fixing the total number of clusters to be 1 in the 
 nucleation rate equation. Besides, this approximation also yields an
approximated equation for the stationary distribution, 
\begin{align}
  P_{s}\sim \frac{\alpha\,\widetilde{F}_0^{1/\alpha}}{\Gamma\left(\frac{1}{\alpha}\right)}g^{1/2}(X)\exp\left(-\beta\Delta F(X)\right).
  \label{eq:stationary-pdf-approximated}
\end{align}

\section{Parametrized profiles\label{sec:parametrizedProfiles}}

The following section is devoted to particularize the expressions derived above to
some specific parametrizations. We will start with the  capillary model,
a crude model where even the smallest clusters {have} zero {interfacial} width. Despite 
being the simplest description of a density fluctuation, the capillary approach
{results} in a robust theory that {captures} the  {most} relevant
aspects of the nucleation process. Thereafter we will test the
effect of considering a finite cluster width under the same circumstances.
When the capillary model is endowed with a surface we call the resulting approach
{the ``extended'' model}.

Our concern is the nucleation of a dense liquid droplet from a weak solution at a given temperature, $T$,
with a finite number of particles (or total mass) $N$ and total volume $V_T$.
The average density of the initial vapor is then given by $\rho_{av}=N/V_T$. In order to
write the vapor density in a simpler way we will refer this quantity to the
coexistence vapor density for an infinite system at the same temperature, which
will be denoted as $\rho_{v}^{\text{coex}}$. The liquid density at the
coexistence, $\rho_{l}^{\text{coex}}$, is then determined by the conditions,
\begin{align}
\omega(\rho_{v}^{\text{coex}}) &= \omega(\rho_{l}^{\text{coex}})\nonumber\\
\omega'(\rho_{v}^{\text{coex}}) &= \omega'(\rho_{l}^{\text{coex}}),
\label{eq:coexistence-conditions}
\end{align}
with $\omega(\rho)=f(\rho)-\mu\rho$ being the free energy per unit volume and
$f(\rho)$ the Helmholtz free energy per unit volume. The ratio
$\rho_{av}/\rho_{v}^{\text{coex}}$
plays a similar role as the supersaturation in ideal systems, thus it will
be referred as the \emph{effective supersaturation}, $S_e$. For that reason, the
initial density will be specified in terms of the effective supersaturation
since we will adopt the convention, $\rho_{av}=\rho_{v}^{\text{coex}}\,S_e$.

\subsection{Modified capillary model\label{sec:capillaryModel}}

The capillary model used in CNT assumes that clusters have no interfacial width and that they emerge with the same properties
as the bulk new phase. In short, that approach can be mathematically expressed as,
\begin{equation}
 \rho(r;R,\rho_0)=\begin{cases}
            \rho_0,\quad r\leq R\\
            \rho_{ext},\quad r>R
           \end{cases}
 \label{eq:capillary-model-general}
\end{equation}
where $R$ is the radius of the cluster, $\rho_0$ is the density
inside the cluster and $\rho_{ext}$ is the value of the
density outside the cluster. In the case of infinite systems, $\rho_{ext}=\rho_{av}$
and $\rho_0$ is  the bulk-liquid density,
which fulfils $\omega'(\rho_{av}) = \omega'(\rho_{l})$. {In contrast, a finite system} 
closed to matter  {exchange cannot} reach  {this} global thermodynamic
equilibrium but {, rather,} a stable state, which will not fulfil the just mentioned
equilibrium condition. {This is because the density of the vapor outside the cluster must drop as the size and density of the cluster grows so as to maintain a fixed number of molecules.} Under these circumstances it seems natural
 to select $\rho_0$ as the stable-state density, $\rho_{st}$,
which yields a minimum of the Helmholtz free energy of the system, $F(\rho(r;R))$
(Eq. \ref{eq:stable-cluster-MCM}).

We have still to express the surrounding density, $\rho_{av}$, as a function of
the {cluster} size. Depending on the total size of the
system, the probability that several fluctuations coexist at the same time will be
negligible or not. In the case in which only one density fluctuation lives
in the system at a time (an assumption always made within CNT), the result of applying the mass conservation law
(Eq. \ref{eq:mass-conservation-law}) gives 
\begin{align}
\rho_{ext}(R,\rho_0)=\frac{\rho_{av}-\delta^3(R)\rho_0}{1-\delta^3(R)}.
\label{eq:rho-ext-capillaryModel}
\end{align}
with $\delta(R)=R/R_T$. This equation explicitly shows that clusters will perturb the surrounding
density as long as the system is small enough. From a straightforward calculation
one observes that $\rho_{ext}(R)\rightarrow\rho_{av}$ as $R_T\rightarrow\infty$,
which is in accordance with the classical description. Combining equations
(\ref{eq:capillary-model-general}) and (\ref{eq:rho-ext-capillaryModel}),
along with $\rho_{0}=\rho_s$, we obtain the modified capillary model (MCM),
\begin{equation}
 \rho(r;R,\rho_0)=\rho_0\,\Theta(R-r)+\rho_{ext}(R,\rho_0)\,\Theta(r-R),
 \label{eq:modified-capillaryModel}
\end{equation}
with $\Theta(x)$ being the Heaviside step function. The Helmholtz free energy
of the system containing a fluctuation, $\beta F(\rho_0,R)$,
will have two contributions. The first one is due to the cluster itself, and it is
postulated to have the common volume-plus-surface structure. The second one is due
to the remaining volume, with density $\rho_{ext}(R,\rho_0)$. Computing the difference
between this energy and that corresponding to the system
with no fluctuation, $\beta F(\rho_{av})$, one gets the work of cluster formation,
\begin{align}
 \Delta \beta F(\rho_0,R) =& \beta F(\rho_0,R)-\beta F(\rho_{av})\nonumber\\
 =&\frac{4\pi}{3}R^{3}\left(\beta f(\rho_{0})-\beta f(\rho_{ext}(R,\rho_{0}))\right)\nonumber\\
 &+4\pi R^{2}\beta \gamma\nonumber\\
 &+\left(\beta f(\rho_{ext}(R,\rho_{0}))-\beta f(\rho_{av})\right) V_T
 \label{eq:W-MCM},
\end{align}
where $\gamma$ is the phenomenological surface tension. The last term will play
a key role in nucleation, since it is related with the presence (or not) of
a global minimum beyond the critical size.

\subsubsection{Critical and stable cluster\label{subsub:criticalAndStableCluster-MCM}}

 {To characterize the critical cluster, we need to minimize the free energy with respect to the cluster's density and radius,} 
\begin{align}
\left(\frac{\partial\beta F(\rho,R)}{\partial R},\frac{\partial\beta F(\rho,R)}{\partial \rho}\right)_{\substack{R=R_{st}\\ \rho=\rho_{st}}}=\mathbf{0}
\label{eq:stable-cluster-MCM}
\end{align}
{where} $R_{st}$ {is the radius of the stable cluster}. {Use of equation (\ref{eq:W-MCM}) then gives}
\begin{equation}
4\pi\,R_{\ast }\left[
\begin{array}{l}
\left( \beta f(\rho_{st})-\beta f(\rho_{ext}(R_{\ast},\rho_{st}))\right)\\
-\left(1-\delta_*^{3}\right)\beta f^{\prime}(\rho_{ext}(R_{\ast},\rho_{st}))
\frac{(\rho_{st}-\rho_{av})}{1-2\delta_*^3+\delta_*^6}
\end{array}\right]=-8\pi\beta \gamma  \label{eq:critical-cluster-MCM}.
\end{equation}
Taking the limit $R_T\rightarrow\infty$, one readily gets
\begin{equation}
R_{\ast }=\frac{-2\beta \gamma}{\left(\beta f(\rho_l)
-\beta f(\rho_{av})\right)-
\beta f^{\prime}(\rho_{av})(\rho_l-\rho_{av})},
\label{eq:critical-radius-MCM}
\end{equation}
that shows the same structure as the equation for open
systems,\cite{book:kashchiev-2000,book:kelton-2010}
\begin{equation}
 R^{\text{CNT}}_\ast =\frac{- 2\beta\gamma}{\beta\omega(\rho_l)-\beta\omega(\rho_{av})}.
\end{equation}
These results clearly show an agreement with those
predicted by CNT, while extending them to situations
where the confinement can play a prominent role, e.g.
inhibiting nucleation for an average density which would proceed
to nucleate in larger systems.

\subsubsection{Cumulative mass and metric\label{subsub:cumulativeMassAndMetric-MCM}}

{The modified capillary model} for the density profile {gives cumulative mass distribution} (\ref{eq:definition-cumulative-mass}),
\begin{align}
 m(r;R) =&\ 4\pi \int_0^{r} \rho(r;R)\ r'^2dr'\nonumber\\
 =&\Theta(R-r)\frac{4\pi}{3}R^3\rho_{0}\nonumber\\
 &+\Theta(r-R)\frac{4\pi}{3}\left(R^3\rho_0 + (r^3-R^3)\rho_{ext}(R)\right)
 \label{eq:mass-MCM}
\end{align}
with $\rho_0=\rho_{st}$. {Using this,} 
the metric can be obtained by employing equation (\ref{eq:mass-MCM})
in (\ref{eq:definition-1dgeneral-gij}) {with the result that}
\begin{align}
 g(R)=g_{1,0,0}(R)+g_{0,1,0}(R)+g_{0,0,1}(R)
 \label{eq:metric-MCM-v1}
\end{align}
with
\begin{align}
g_{1,0,0}(R)=&\frac{4\pi R^{3}}{\rho_{ext}(R)}
  \left(\rho_{ext}(R)-\rho_{0}\right)^{2}\left(1-\delta(R)\right)
  \label{eq:metric-cap-comps}\\
g_{0,1,0}(R) =&-\frac{4\pi R^{2}}{3\rho_{ext}(R)}
  \left(\rho_{ext}(R)-\rho_{0}\right)\nonumber\\
  &\times  \frac{(R_{T}-R)^{2}(R_{T}+2R)}{R_{T}}
  \left(\frac{\partial\rho_{ext}(R)}{\partial R}\right)
  \nonumber \\
g_{0,0,1}(R) =&\frac{4\pi R_{T}^{3}}{45}
  \left(1+3\delta(R) +6\delta^{2}(R)+5\delta^{3}(R)\right)\nonumber\\
  &\times\frac{(R_{T}-R)^{3}}{R_{T}}
  \left(\frac{\partial\rho_{ext}(R)}{\partial R}\right)^{2}\nonumber
\end{align}
It is easy to check that $g_{0,1,0}(R)$ and $g_{0,0,1}(R)$ tend to $1/R_T$ when
$R\ll R_T$, so they represent small corrections to the first term $g_{1,0,0}(R)$.
 {Thus}, we recover the metric derived for infinite
systems\cite{article:lutsko-duran-2013} when that limit is considered.
That fact leads us to rewrite equation (\ref{eq:metric-MCM-v1}) as,
\begin{align}
 g(R)=&\ g_{1,0,0}(R)\left(1+\frac{g_{0,1,0}(R)+g_{0,0,1}(R)}{g_{1,0,0}(R)}\right)\nonumber\\
 =&\ g_{1,0,0}(R)\ \chi(R)
 \label{eq:metric-MCM-v2}
\end{align}
As we discussed in section \ref{sec:theory}, the inverse of the metric plays a similar
role as the monomer attachment rate in the Zeldovich-Frenkel
equation when the order parameter is
the number of particles. To test this fact we perform the change of variable
\begin{equation}
 \Delta N = \frac{4\pi}{3}R^3(\rho_0-\rho_{ext}(R)).
 \label{eq:deltaN-MCM}
\end{equation}
The metric is easily translated to the new variable,
\begin{align}
 g^{-1}(\Delta N)=&\left(\frac{d\Delta N}{dR}\right)^2g^{-1}(R(\Delta N))\nonumber\\
 =&\ \zeta(\Delta N) 4\pi R(\Delta N) \rho_{ext}(R(\Delta N))
\end{align}
with
\begin{equation}
\zeta (\Delta N)=\frac{\left(1-\frac{\rho_{0}-\rho_{av}}{\rho_{0}-\rho_{ext}(R(\Delta N))}
\delta^{3}(\Delta N)\right)^{2}}{%
1-\delta (\Delta N)}\chi^{-1}(R(\Delta N))  \label{eq:sticking-coeff}
\end{equation}
It turns out that $f(\Delta N)=Dg^{-1}(\Delta N)$ {has} essentially the same structure as
the usual result for the monomer attachment rate within
the context of diffusion-limited nucleation: indeed the first converges to the second
when $R_T\rightarrow\infty$. Note that here $\zeta(\Delta N)$ would be
the counterpart of the phenomenological sticking coefficient.

\subsubsection{Expansion of the canonical variable \label{subsub:canonicalVariable-MCM}}

With the aid of the expression of the metric we can look for the canonical variable
defined by (\ref{eq:canonical-transformation}), 
\begin{align}
 Y(R')=\int_0^{R'}\sqrt{\frac{4\pi\,R^3(\rho_{ext}(R)-\rho_0)^2}{\rho_{ext}(R)}
 (1-\delta(R))\chi(R) } dR.
 \label{eq:canonical-Y-MCM}
\end{align}
However, the canonical variable is not an elementary function of $R$ due to the complexity of the integrand.
Fortunately, the practical interest on this variable resides in obtaining a first-order approximation
of the work of cluster formation and the number of particles inside a cluster in the case of small clusters.
Under such circumstances one can consider $\delta(R)\sim 0$ as a good approximation and, therefore,
$g(R)\sim g_{1,0,0}(R)$. Thus,
\begin{align}
 Y(R) \sim &\ \frac{2}{5}
      \left(\frac{4\pi(\rho_{0}-\rho_{av})}{\rho_{av}}\right)^{1/2}
      R^{5/2}\label{eq:can-var} \\
 R(Y) \sim &\ \left(\frac{5}{2}\left(
      \frac{4\pi(\rho_{0}-\rho_{av})}{\rho_{av}}
      \right)^{-1/2}\right)^{2/5}Y^{2/5}\nonumber
\end{align}
so that,
\begin{align}
\Delta\beta F(Y) \sim &\ 4\pi \beta\gamma R^{2}(Y) \nonumber\\
          \sim&\ 4\pi\beta\gamma\left(\frac{5}{2}
                  \left(\frac{\rho_{av}}{4\pi(\rho_{0}-\rho_{av})}\right)^{1/2}
                  \right)^{4/5} Y^{4/5}
                  \label{eq:F-expansion-MCM}
\end{align}
and,
\begin{align}
\Delta N \sim &\frac{4\pi}{3}\left(\rho_{0}-\rho_{av}\right)
       \left(\frac{5}{2}\left(\frac{\rho_{av}}{4\pi(\rho_{0}-\rho_{av})}
       \right)^{1/2}\right)^{6/5}Y^{6/5}
       \label{eq:DeltaN-expansion-MCM}
\end{align}
These expressions are used in the calculations of the nucleation rate
in order to make simpler the integrals involved (Eq. \ref{eq:escape-rate-general}), 
with $\alpha=\frac{4}{5}$ and where
\begin{equation}
 \widetilde{F}_0=4\pi\beta\gamma\left(\frac{5}{2}
                  \left(\frac{\rho_{av}}{4\pi(\rho_{0}-\rho_{av})}\right)^{1/2}
                  \right)^{4/5}
\end{equation}

\subsubsection{The stochastic differential equation\label{subsub:SDE-MCM}}

{The SDE now becomes}
\begin{align}
 \frac{dR}{dt}=&-Dg^{-1}(R)\frac{\partial}{\partial R}\left(\Delta\beta F(R)+\frac{1}{2}\ln g(R)\right)\nonumber\\
 &+\sqrt{2D\,g^{-1}(R)}\,\xi(t)\nonumber.
 \label{eq:sde-R-MCM}
\end{align}
When the cluster and system are large enough the SDE {converges to}
that derived by \citeauthor{article:lutsko-duran-2013},\cite{article:lutsko-duran-2013}
which yields the classical result $R\sim t^{1/2}$ when the higher order terms in $R^{-1}$
are neglected.\cite{book:saito-1998} {In contrast,} in confined systems the result
{is very different} when the cluster is  large compared to the total volume. In that situation,
the mass conservation law {does not allow} the cluster {the cluster to grow indefinitely}, as it does in CNT and dCNT.
Indeed, clusters will not be able to grow beyond the stable size determined by
equation (\ref{eq:stable-cluster-MCM}). Accordingly, the modified capillary model is
able to reproduce the slow down of the growth rate of post-critical clusters {expected in a confined system}, unlike
the classical theory.

\subsection{Extended model\label{sec:extendedModels}}
\subsubsection{The profile and the metric\label{subsub:profileAndMetric-EM}}

One of the most  {obvious} deficiencies of the capillary model is the
zero-thickness  {interface} assumed for clusters, even for the smallest ones
where most of molecules will lie on the cluster surface. To circumvent such a
limitation, piecewise-linear profiles (PLP) have been used in previous works,
\cite{article:lutsko-2011-b,article:lutsko-2012-dtn,article:lutsko-duran-2013}
allowing thus a smooth transition from the inner to the outer density value. {We use the same idea to extend the MCM profile as,}
\begin{equation}
 \rho(r)=\begin{cases}
          \hfill{}\rho_0, & r<R-w\\
          \hfill{}\rho_0-(\rho_0-\rho_{ext}(R))\frac{r-(R-w)}{w},&R-w<r<R\\
          \hfill{}\rho_{ext}(R),&R<r
          \end{cases}
 \label{eq:extendedModel-general}
\end{equation}
where the density out of the cluster is determined by the mass conservation law,
\begin{align}
\rho_{ext}(R)=&\frac{\rho_{av}-\left(\delta^3(R)-\psi(R;w)\right)\rho_0}{%
		      1-\left(\delta^3(R)-\psi(R;w)\right)}\nonumber \\
\psi(R;w)=&\frac{4\pi}{w V_T}
	  \left(
	  \begin{array}{l}
	  \frac{R^4-(\max(R-w),0)^4}{4}\\
	  \qquad+\frac{(w-R)\left(R^3-(\max(R-w,0))^3\right)}{3}
	  \end{array}
	  \right)
\label{eq:rho-ext-EM}
\end{align}
so that $m(R_T)/V_T=\rho_{av}$. The parameters $\rho_0$ and $w$ have to be {fixed} according
to some reasonable physical criterion. In order to be consistent with the previous section,
the inner density will be set to minimize the free energy of the stable
cluster. Following the same reasoning, it seems natural to set the width parameter as that
fulfilling the same rule. To this end we need to construct the free energy model for the PLP
(Eqs. \ref{eq:extendedModel-general} and \ref{eq:rho-ext-EM}) and solve the 3-dimensional
root-finding problem,
\begin{align}
 \left(
    \frac{\partial \beta F(\mathbf{X})}{\partial X_j}
 \right)_{\mathbf{X}=\{R_{st},\rho_{st},w_{st}\}}=0
 \label{eq:stable-cluster-EM}
\end{align}
These {do not permit} an exact solution {and so will be solved numerically.}

\subsubsection{Free energy model\label{subsub:freeEnegyModel-EM}}

The aim of this work is ultimately make a connection with the calculations already performed 
for infinite systems. Thus, the model for the free energy in the PLP approach will be constructed
based on a simple\cite{article:lutsko-2011-c}
\begin{equation}
 F[\rho]=\int\left(f(\rho(\mathbf{r})+\frac{1}{2}K
 		\left(\nabla \rho(\mathbf{r})\right)^2\right)d\mathbf{r}
 \label{eq:squared-gradient}
\end{equation}
where $K$ is the squared-gradient coefficient that will be estimated by using
the results of Ref. \onlinecite{article:lutsko-2011-c}, and the Helmholtz free 
energy per unit volume can be calculated based on a pair potential using 
thermodynamic perturbation theory or liquid state integral equation methods.
Substituting the PLP into equation (\ref{eq:squared-gradient}) yields, 
\begin{widetext}
\begin{align}
\beta\Delta F(R;w)=&\frac{4\pi}{3}\left(\max(R-w,0)\right)^3\beta\Delta f(\rho_0)
+\left(1-\delta^3(R)\right)\,\beta\Delta f(\rho_{ext}(R))\,V_T  \notag \\
&+\int_{\max(R-w,0)}^{R}4\pi\,r^2\,\beta\Delta
f\left(\rho_0-(\rho_0-\rho_{ext}(R))\frac{r-R+w}{w}\right)dr  \notag \\
&+\frac{\beta K}{2}\frac{4\pi}{3}\left(R^3-\max(R-w,0)^3\right)\left(\frac{%
\rho_0-\rho_{ext}(R)}{w}\right)^2
\label{eq:W-EM-squaredGradient} 
\end{align}
\end{widetext}
{which} equation has exactly the same structure {as} that derived for infinite systems
except for the second {term} which accounts for the confinement.

\section{Results and comparisons\label{sec:resultsAndComparisons}}

The theory previously presented was evaluated by considering a model of globular proteins, as {was previously} done in the case of infinite systems. Thus, the solvent was approximated by considering 
Brownian dynamics of the solute molecules which simultaneously experience an effective pair potential that we 
assumed to be the ten Wolde-Frenkel potential,\cite{article:tenwolde-frenkel-1997}
\begin{align}
 v(r)=\begin{cases}
              \infty, & r\leq\sigma\\
              \frac{4\epsilon}{\alpha^2}\left(\left(\frac{1}{\left(\frac{r}{\sigma}\right)^2-1}\right)-
              \left(\frac{1}{\left(\frac{r}{\sigma}\right)^2-1}\right)\right), & r\geq\sigma\\
             \end{cases}
\end{align}
with $\alpha=50$ which is then cutoff at $r_c=2.5\,\sigma$ and shifted so that $v(r_c)=0$. 
With the aim to compare the results obtained with the present theory with those reached
for infinite systems we fixed the temperature at $k_BT=0.375\,\epsilon$. 
{The} free energy density $f(\rho )$ was computed {using} thermodynamic 
perturbation theory. Finally, the squared-gradient coefficient was calculated making 
use of the results in Ref. \onlinecite{article:lutsko-2011-c}, i.e.
\begin{align}
\beta K\simeq -\frac{2\pi }{45}d^{5}\beta v(r)+\int_{d}^{\infty }\left(
2d^{2}-5r^{2}\right) v(r)r^{2}dr  \label{eq:K-calculation-app}
\end{align}
with $d$ being the effective hard-sphere diameter. Under these conditions, it was 
shown that the squared-gradient coefficient is $\beta K=1.80322\,\sigma^5$. Finally, 
the CNT value for the surface tension was computed by using the following expression 
for a planar interface,\cite{article:lutsko-duran-2013}
\begin{equation}
\gamma_{\text{CNT}}=(\rho_0^\text{coex}-\rho_{av}^\text{coex}) \sqrt{2K\overline{\omega}_0^{\text{coex}}},
\end{equation}
with 
\begin{equation}
 \overline{\omega}_0^\text{coex}=\frac{1}{(\rho_0^{\text{coex}}-\rho_{av}^{\text{coex}})}\int_{\rho_{av}^{\text{coex}}}^{\rho_0^{\text{coex}}}(\omega(x)-\omega(\rho_{av}^{\text{coex}}))dx.
\end{equation}

\subsection{Work of cluster formation}
{The} energy barrier for cluster formation {is a key quantity in }  nucleation theories as well as 
the comparison of its value for different average densities, or supersaturation values under CNT conditions.
In  {order to make} contact with the results obtained for infinite systems, we {use as independent variable the } effective 
supersaturation, $S_e$, which is {the average density divided by} the coexistence density 
(for infinite systems) at the given temperature. {W}e evaluated the free energy models proposed in 
section \ref{sec:parametrizedProfiles} for effective supersaturations from $S=1.125$ to
$S=2.5$ {thus} covering a wide range of critical sizes, from very large to very small.

\begin{figure}[t]
\centering{}
  \includegraphics[width=0.49\textwidth]{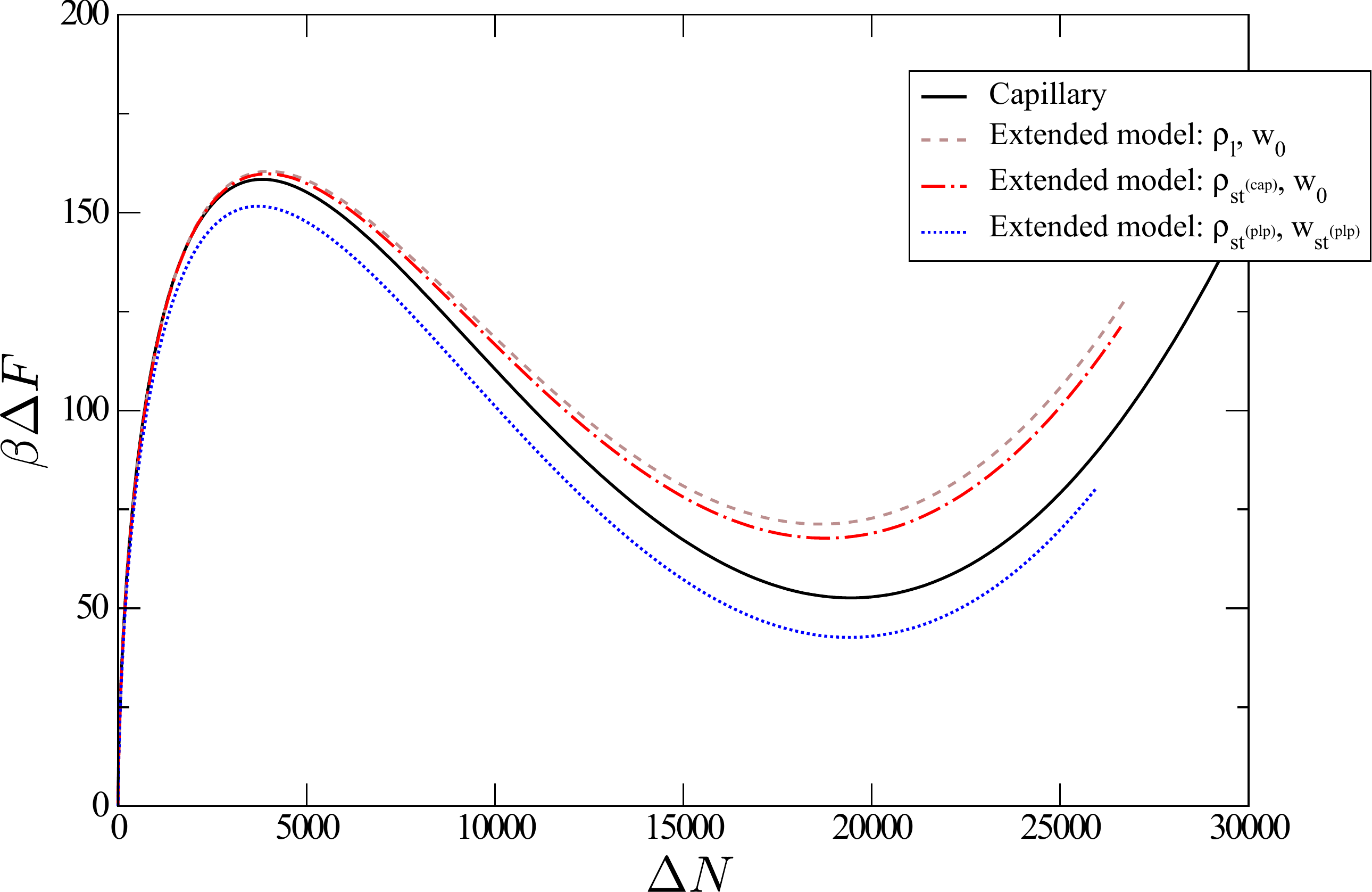} 
\caption{\label{fig:F-1125} The free energy of cluster formation as a function of 
number of molecules inside the cluster at $S_e=1.125$ in a confined system 
of total volume, $V_T=4.5\times 10^6\sigma^{-3}$, using the modified capillary model 
with $\gamma$ being that calculated from infinite systems, and the extended model, 
with which we tested different combinations of the characteristic parameters $\rho_0$ and $w$.
This graph shows the fact that the liquid phase is not stable so that nucleation
will not proceed.}
\end{figure}

\begin{figure*}[t]
\begin{center}
 \includegraphics[width=0.35\textwidth]{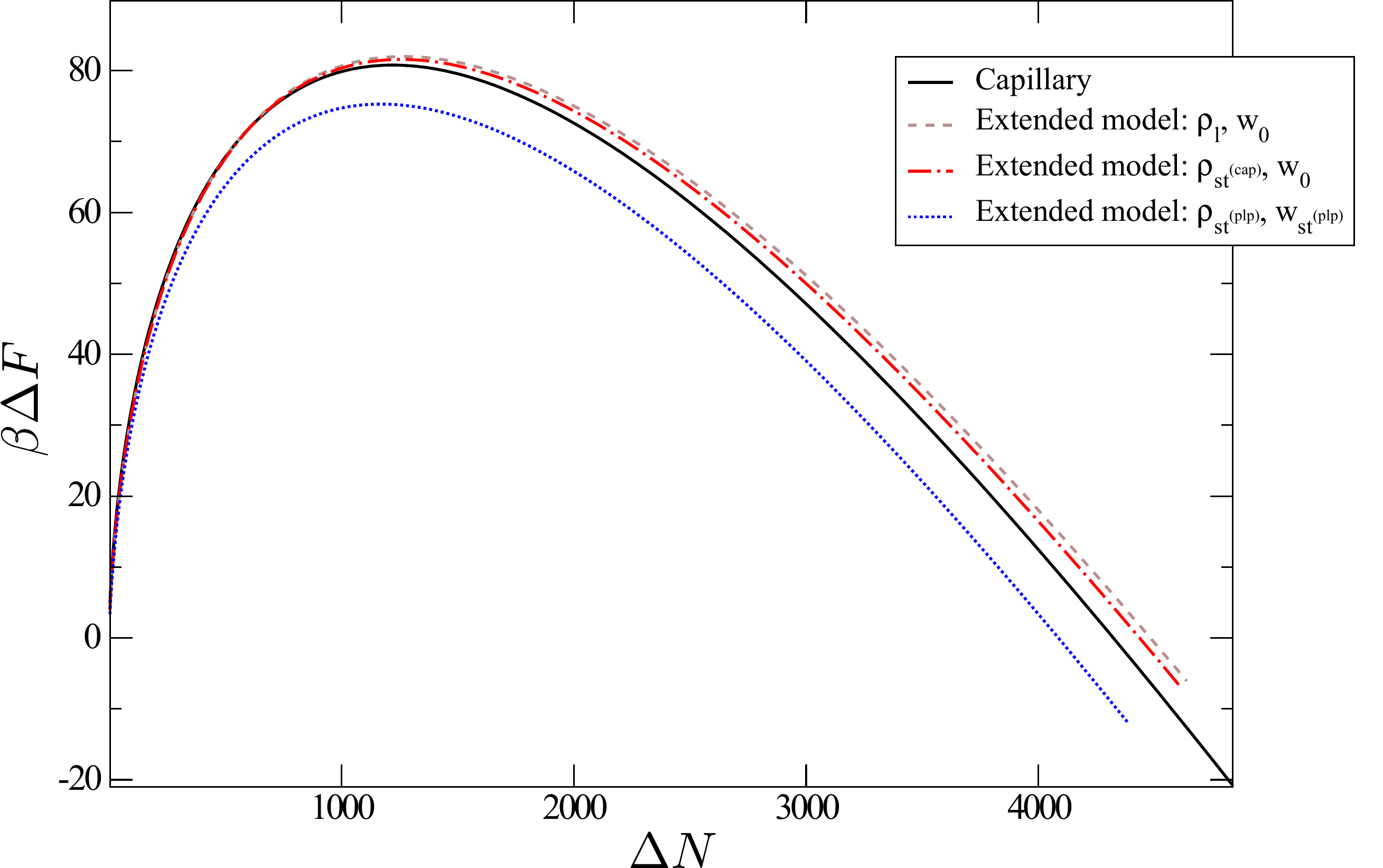} 
 \includegraphics[width=0.35\textwidth ]{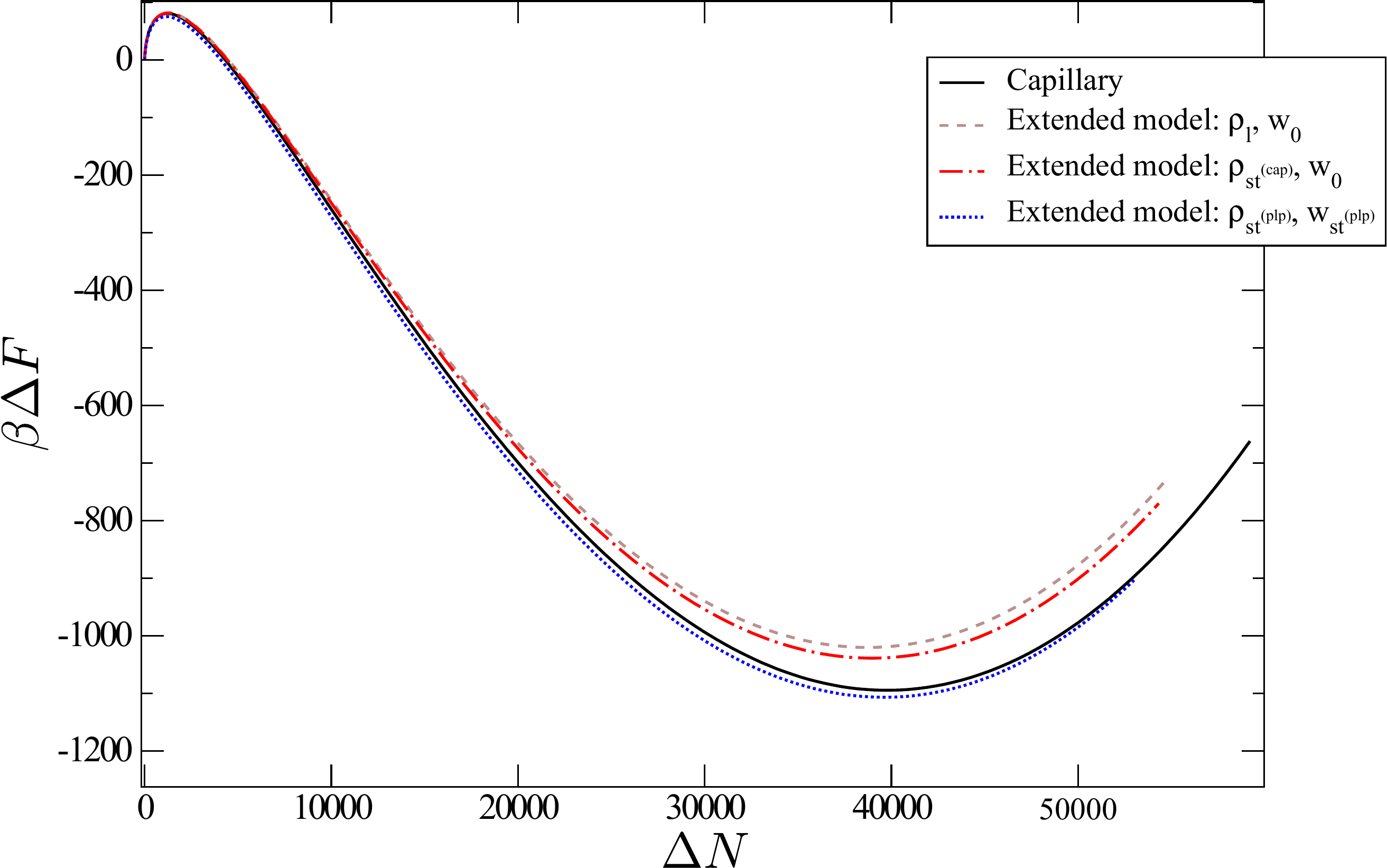}\\
 \vspace{0.15cm}
 \noindent{}\includegraphics[width=0.35\textwidth ]{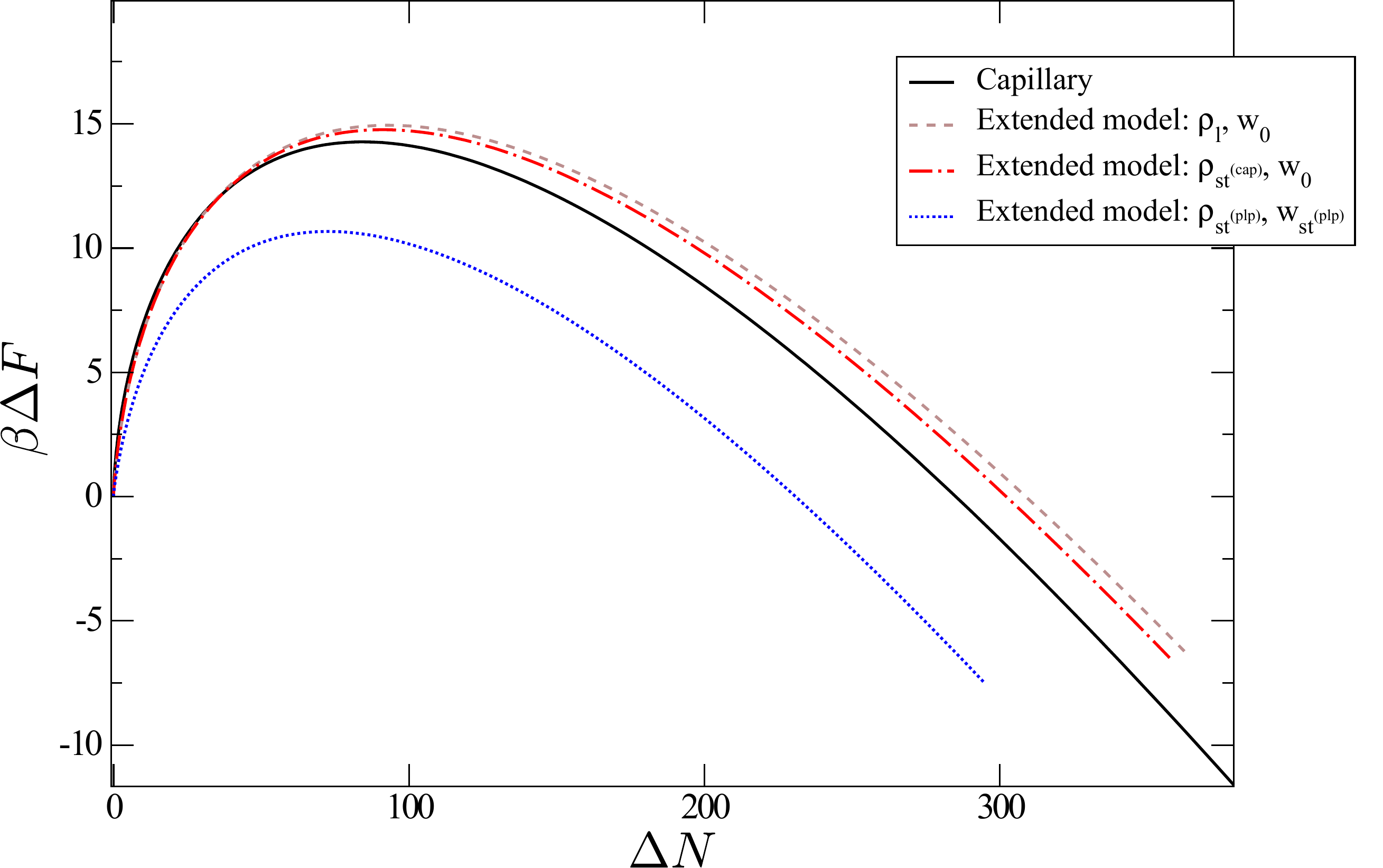} 
 \includegraphics[width=0.35\textwidth ]{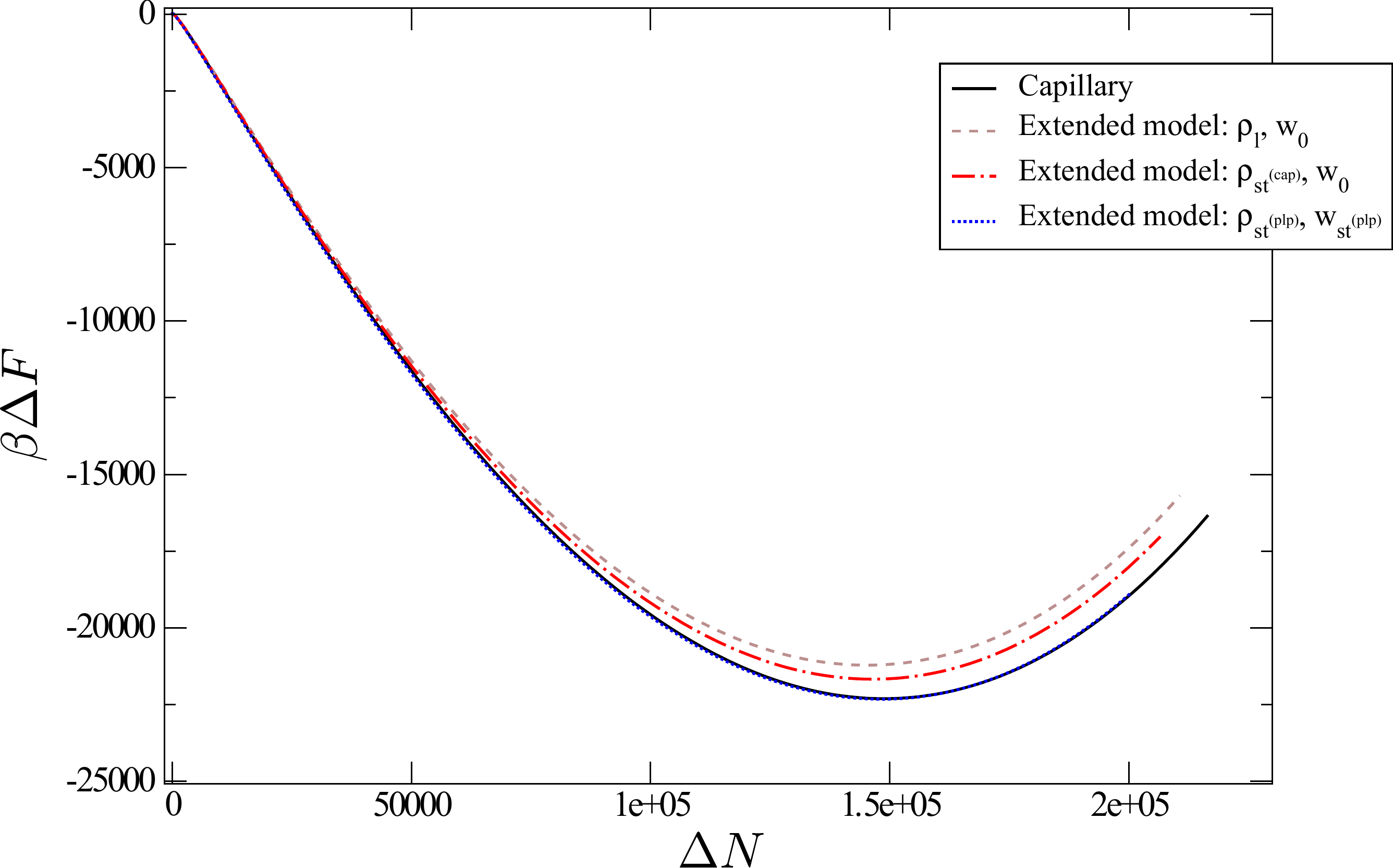}\\
 \vspace{0.15cm}
 \noindent{}\includegraphics[width=0.35\textwidth ]{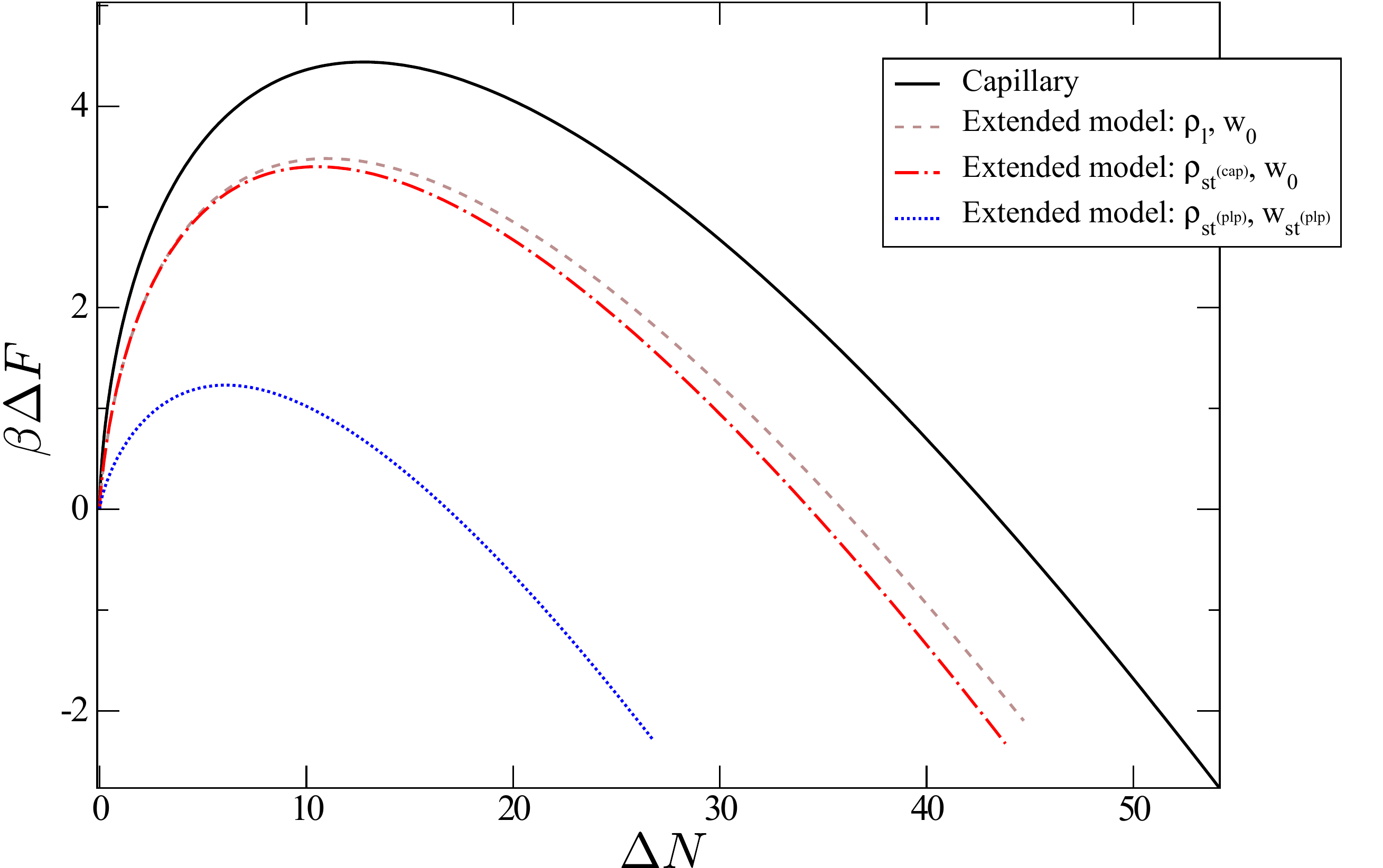} 
 \includegraphics[width=0.35\textwidth ]{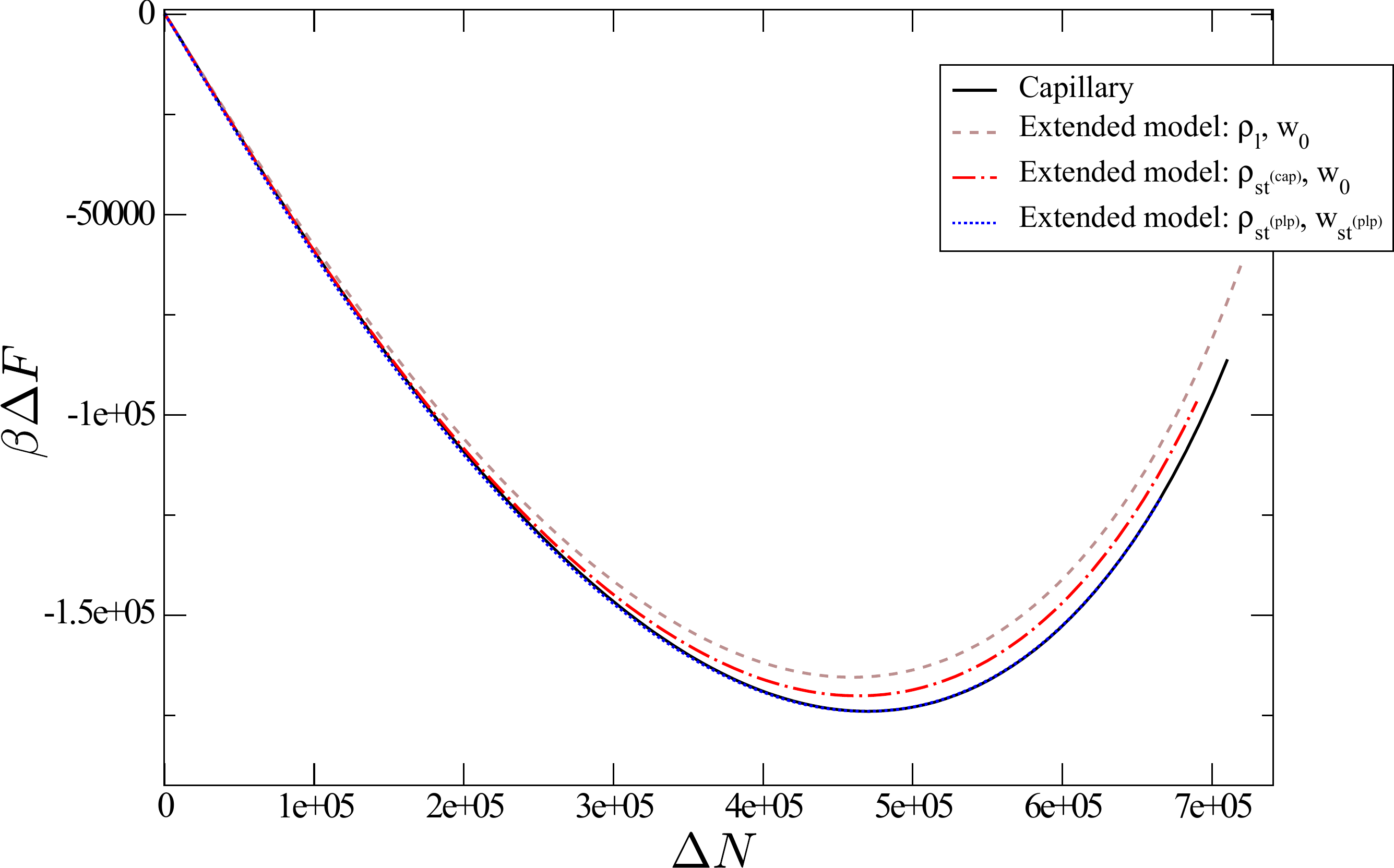}
 \caption{The Helmholtz free energy as a function of cluster size, $\Delta N$, at $S=1.175,\,1.5\ $and $2.5$ for different cluster models. The left column are zooms of the figures on the right column about the critical size at each supersaturation. The right column shows the existence of a stable size behind which the energy of formation rockets. The total volume is again $V_T=4.5\times10^{6}\sigma^{-3}$.}
 \label{fig:F-all}
\end{center}
\end{figure*}

The work of cluster formation was evaluated by using equations (\ref{eq:W-MCM}) and 
(\ref{eq:W-EM-squaredGradient}) for the modified capillary and extended models respectively.  Concerning the MCM,
we fixed the surface tension to equal the CNT value calculated in the previous study for infinite systems,\cite{article:lutsko-duran-2013}
i.e. $\gamma=\gamma_{\text{CNT}}$. However, the inner density $\rho_0$ was {adjusted so as } to
minimize the free energy of the stable cluster, $\rho_{{st}^{(cap)}}$, 
unlike the classical capillary model where the inner density is set to be the 
that of the new phase, $\rho_l$. As for the extended model considering the PLP,
we studied several possibilities to choose the characteristic parameters so that
we can {see} more easily the effects of: the confinement, the interior density 
and the surface width. Thus, we tested three different combinations of the values 
$\rho_0$ and $w$:
\begin{enumerate}
 \item[a)] Set the density $\rho_0=\rho_l$ and $w=w_0$ (from dCNT),
 \item[b)] Set the density $\rho_0=\rho_{{st}^{(cap)}}$  and $w=w_0$,
 \item[c)] Look for the pair $(\rho_{{st}^{(plp)}},w_{{st}^{(plp)}})$ to minimize the 
 free energy (Eq. \ref{eq:W-EM-squaredGradient}) of the stable cluster.
\end{enumerate}
{F}igures \ref{fig:F-1125} and \ref{fig:F-all} show
the free energy landscapes at $S_e=1.125$ and $S_e=1.175,1.5, 2.5$, respectively. The reason why
the supersaturation values {was} divided into subsets is to highlight the fact that nucleation is 
inhibited in the first case while it still occurs in the other, as {is} obvious {from} these figures. 
On the one hand, in both cases we can observe the most important effect 
of considering confinement  {which} is the emergence of a local (stable or metastable) minimum
beyond the critical size as a result of finite mass. This is a new property which {has no counterpart for an infinite system.} 
Depending on the total amount of material, such a minimum will become metastable 
(Fig. \ref{fig:F-1125}) or stable (Fig. \ref{fig:F-all}). On account of this fact 
a new effect arises, namely the control on the nucleation rate and the nucleation itself
as a function of the total volume. Indeed, with the volume previously specified 
at $S_e=1.125$ no nucleation event will occur, given that the liquid (supposedly the
new phase) is not stable any more. On the other hand, it is clear from those figures 
that the capillary model with a fixed $\gamma$ produces results close to those
obtained with the extended models, at least up to $S_e=1.5$. {In addition}, we observe how
the interface width plays a key role in the finite-width models lowering
the energy of both the critical and the stable cluster, 
since $\rho_{st^{(cap)}}\simeq\rho_{st^{(plp)}}\simeq \rho_l$ (see Table \ref{tab:j}). Indeed, what we found is that
the width value which minimizes the stable-cluster energy is about twice the value 
$w_{0}$. {There} is also observed 
a great similarity of these results with respect to those for the infinite case, 
if we only pay attention on the left column of Fig. \ref{fig:F-all}. Finally,
in view of these results an interesting conclusion can be drawn in terms of
experimental setups. The control on the total volume enables to 
modulate the stability of a given phase. This is an interesting result for 
crystallization experiments in small volumes (e.g. microfluidics), 
since it would imply that the effective solubility curve 
could be controlled at will.

\begin{figure*}
\begin{center}
 \includegraphics[width=0.325\textwidth]{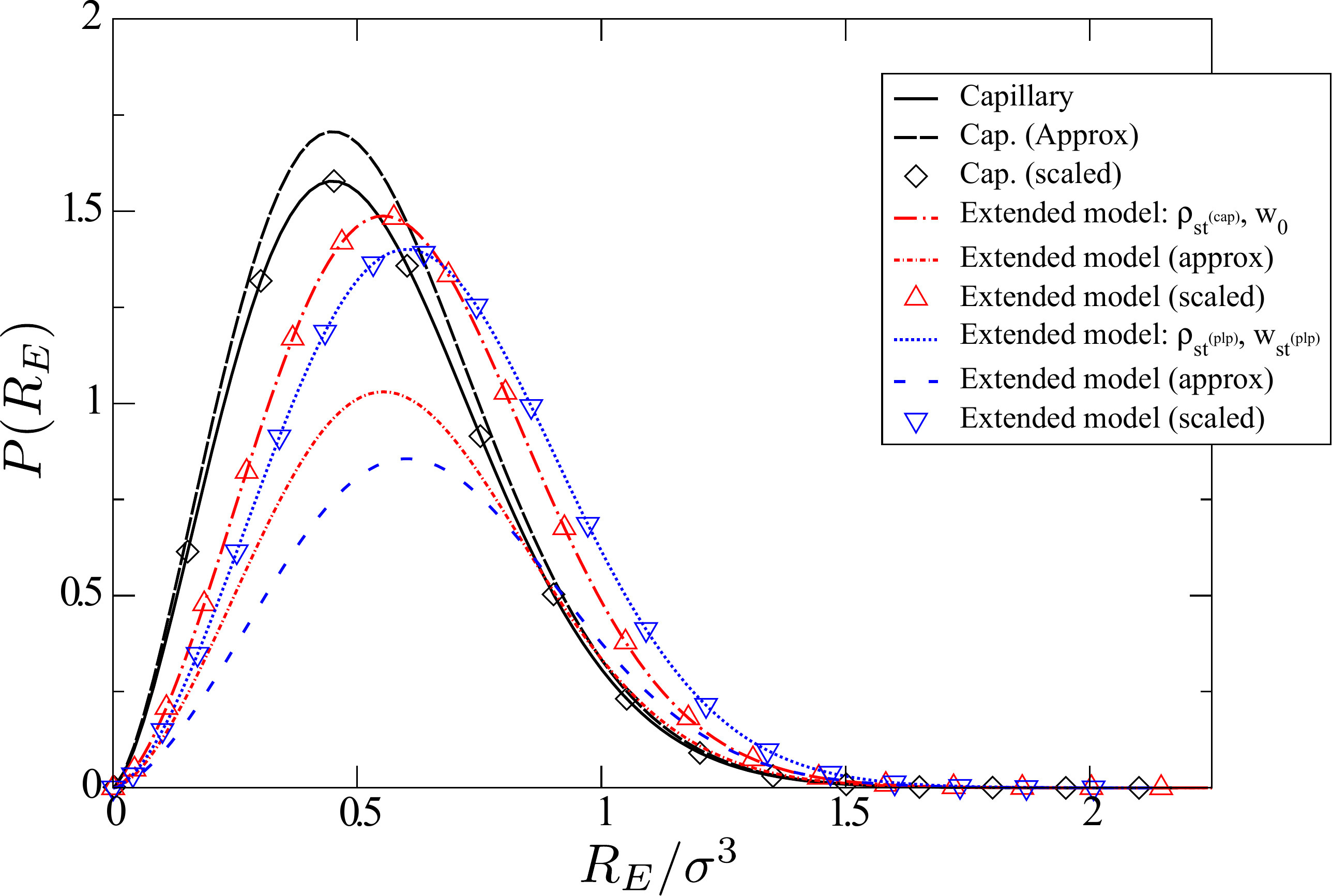} 
 \includegraphics[width=0.325\textwidth]{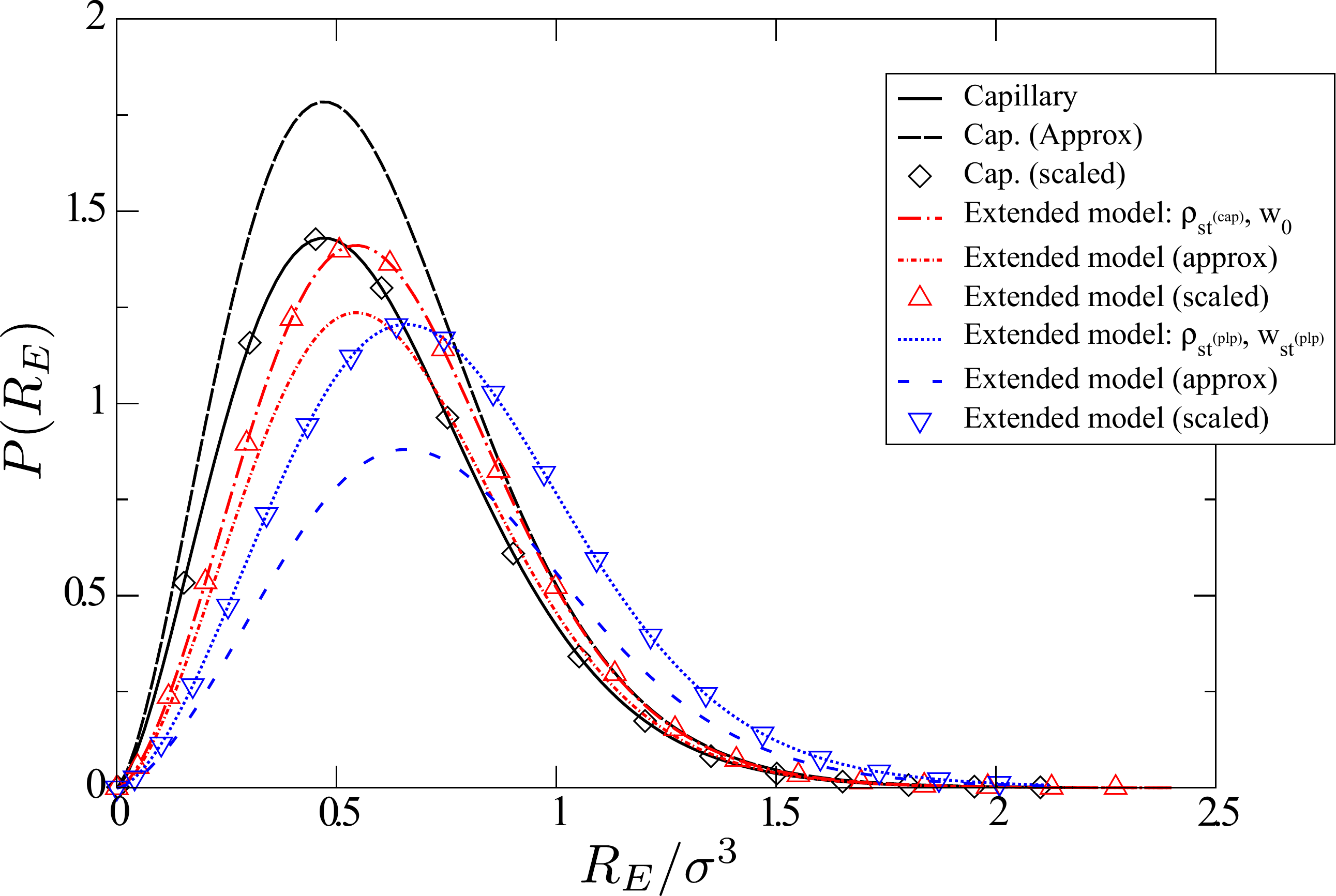} 
 \includegraphics[width=0.325\textwidth ]{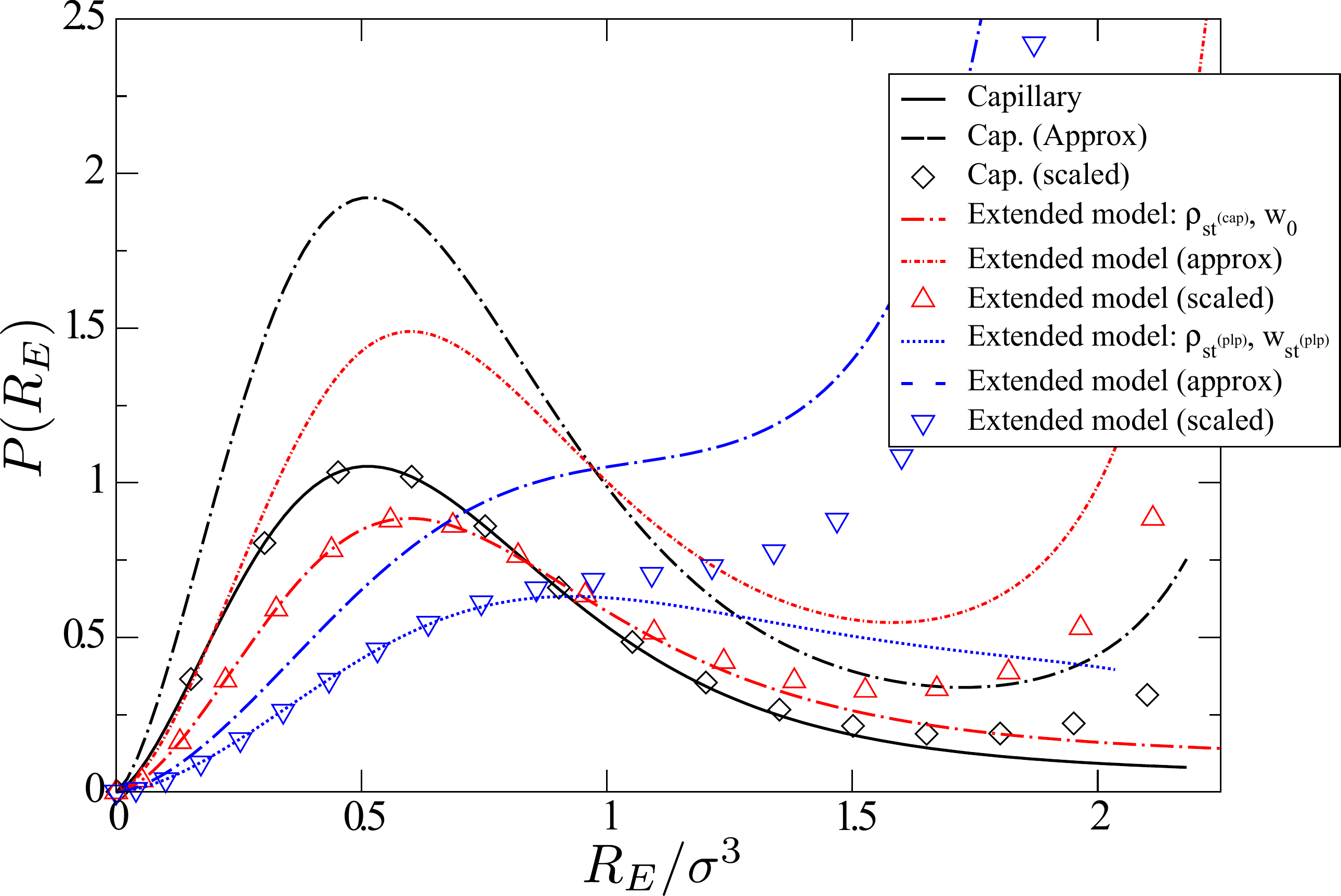}
 \caption{The stationary size distribution for the supersaturation values under which
 nucleation can proceed, $S_e=1.175$ (left panel), $S_e=1.5$ (center panel) 
 and $S_e=2.5$ (right panel), with $V_T=4.5\times10^{6}\sigma^{-3}$ and $R_+=1.5R_*$.}
 \label{fig:Pst}
\end{center}
\end{figure*}

\subsection{The stationary distribution}

A straightforward connection with  experimental measurements can be {made} via
the PDF which is essentially the quantity obtained by techniques like 
\emph{dynamic light scattering} (DLS).\cite{book:berne-2000-dls} 
Thus, the stationary PDF offers us another way to test the theories presented 
above. {In addition},  this quantity is required 
both in its exact (Eq. \ref{eq:steady-state-final-pdf}) and approximated 
(Eq. \ref{eq:stationary-pdf-approximated}) {form so as to determine the nucleation rate ans so}  we need to test its validity. 
In order to do that, we have to compute the PDF for the different models 
in terms of a common variable since $R$ does not mean the same thing in 
both of them, as we already pointed out. Thus, the calculations will be
performed using the equimolar radius, $R_E$, {which} requires the
transformation,
\begin{align}
 \overline{P}(R_E)=P(R)\frac{dR}{dR_E},
\end{align}
with $R_E$ being equivalent to $R$ for the MCM or being given by equation (\ref{eq:equimolar-R-EM})
for the PLP.

The stationary size distributions are {displayed} in figure \ref{fig:Pst} showing  {good} 
agreement with the results for the infinite case.  The {shape of the } PDF is 
faithfully reproduced  by the approximated equation (Eq. \ref{eq:PDF-equilibrium}), at least for 
the lower effective supersaturations (left and center panels).  {However},
the normalization is not equally well estimated, which is a result
of the rapid change of the free energy with the cluster size for small clusters.
Secondly, while for the MCM the approximation still remains being a good estimation
for the highest density ($S_e=2.5$), 
a {significant} error arises for the extended models. The worse 
result lies on the extended model with a minimized stable cluster due to the 
fact that the system is {in}  the pseudospinodal 
region,\cite{article:xu-ting-kusaka-wang-2014-reviewNucleation} i.e. $\beta\Delta F_*\sim 1$, 
so that {the assumption} that small sizes govern the integral result is quite crude.  
Indeed, for these density values one would expect that cluster-cluster interactions
play a key role thus violating the hypotheses assumed to make these calculations,
as was noticed for infinite systems. Notwithstanding, {we conclude that} the capillary model exhibits a
surprising ability to {capture} the main properties of nucleation even for 
finite systems.

\subsection{Nucleation rates}

\begin{figure}[t!]
\begin{center}
 \includegraphics[width=0.48\textwidth]{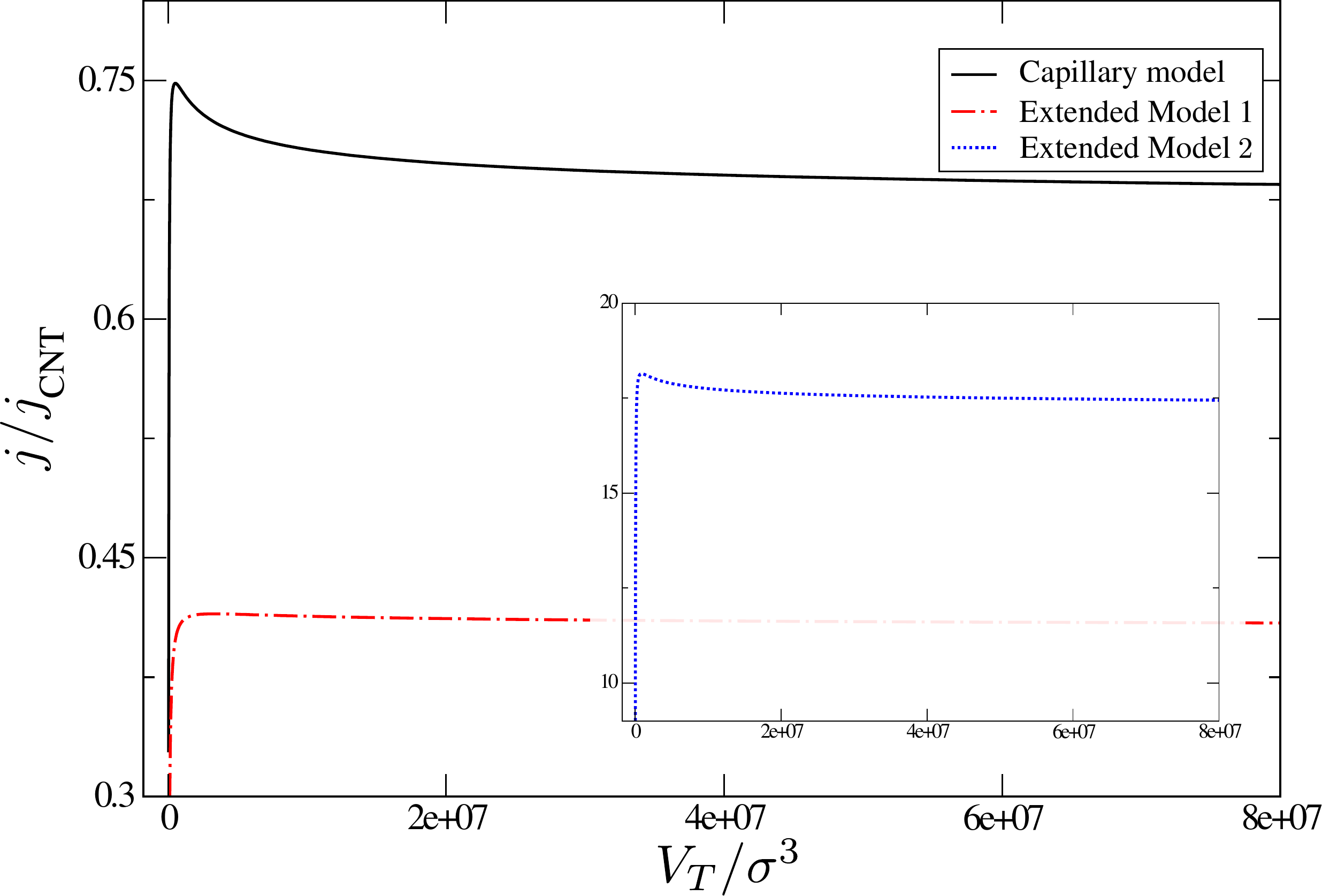}
 \caption{Nucleation rates as functions of the total volume at $S_e=1.5$. The extended models
 considered here are: 1) considering $\rho_0=\rho_{st^{(cap)}}$ and $w=w_0$, and
 2) taking $\rho_0=\rho_{st^{(plp)}}$ and $w=w_{st^{(plp)}}$, i.e. those values which
 minimize the stable-cluster energy. It is observed how the confinement takes
 effect in a very narrow region, resulting in a maximum nucleation rate before
 inhibiting the process.}
  \label{fig:j_VT_15}
\end{center}
\end{figure}

{We end by comparing} the nucleation (escape) rates 
in the different {models} previously introduced {as shown} in Table \ref{tab:j}. It is  {apparent} that for the lower densities the 
nucleation rates are much lower for the extended models with $w$
taken from dCNT calculations than for the capillary model, which is essentially due to the higher 
energy barrier associated with both of them. {On the other hand}, one observes the opposite situation
when the extended model with a minimized stable cluster is considered, since the
energy barrier is lower than that of the MCM (see Fig. \ref{fig:F-all}).  
For the other cases the capillary approximation yields similar results 
to the extended models and to the CNT predictions. 
{Next, we consider the variation of the nucleation rate as a function of volume.} For the sake of simplicity, since a similar
result in shape is obtained for each density we selected $S_e=1.5$. This 
calculation is shown in Fig. \ref{fig:j_VT_15}. A surprising effect is observed
near the zero-rate zone, the nucleation rate exhibits a maximum for very small
volumes and after that relaxes quickly to a steady value, which is nearly the one
presented in Table \ref{tab:j}. This is the result of a competition between two 
effects. On the one hand, the inner density of the cluster decreases with increasing total radius
so that the bulk free energy increases. On the other hand, the free energy associated
to the zone outside the cluster decreases when the total volume increases. It is therefore such a competition
which causes a minimum in free-energy barrier and, hence, the maximum in nucleation rate.
Before that maximum, the nucleation rate passes from
being zero to non-zero in a very narrow region. From this result we can draw the 
conclusion that confined systems could be pretty well approximated by the infinite-system
predictions, unless the volume under consideration {is very} close to the {minimum} volume
{for} nucleation.

\section{Conclusions\label{sec:conclusions}}

In this work,  {a recent reformulation of}  classical nucleation theory\cite{article:lutsko-duran-2013}
 {has been extended}  
to consider finite systems. The {motivation for} making such {an} effort {arises from}  the 
explosion of interest  {in} nucleation process by using new techniques, such as microfluidics,
where the hypotheses made by CNT are probably far from reality. Given that the 
dynamical reformulation of CNT was founded in a more fundamental framework, 
it was relatively easy to modify its derivation to take into account the mass-conservation law 
along with a finite volume and {to} go beyond the initial scope of CNT. With this goal
attained, general expressions for both the stationary distribution function and
the nucleation rate were obtained. Those were ultimately used with two different parametrized
density profiles, a modified version of the capillary model to consider mass conservation
and a piecewise-linear profile. Thus, the results obtained thereby allow to make a {direct} 
comparison to those performed for infinite systems.

The main conclusion we can draw from this study is that the
nucleation rate can be somehow enhanced in a confined system. 
However, confinement affects in practice a very 
narrow range of volumes {which is also} why {CNT} produce{s} good estimates. 
Surprisingly, the different profiles
proposed here gave similar results where the main difference between them lies on
the free energy barrier, as it also does for infinite systems. That said, it seems to
us that the most natural way to further develop dCNT would be allowing the inner density
to freely vary within the capillary model, which {seems a good} balance between
being simple and accurate.

The nucleation rates were calculated in terms of the mean first-passage 
time,\cite{article:hanggi-talkner-borkovec-1990,article:wedekind-strey-reguera-2007,article:lundrigan-2009}
which has been a widely used {approach} in this field. These calculations involved 
similar ingredients to those required to compute the stationary distribution function.
The latter was evaluated numerically (Eq. \ref{eq:steady-state-final-pdf}) 
and by using its approximated version (Eq. \ref{eq:stationary-pdf-approximated}).
 A good agreement between exact and approximated expressions were found for 
low and intermediate densities while for higher values the approximation 
became less accurate. Therefore, the same can be observed in the nucleation
rates in Table \ref{tab:j}. However, the fact that high densities yield 
worse approximations is not a key problem since certainly in such a regime
the hypothesis of non-interacting clusters will be unlikely valid any more.

Finally, the volume of the system under study was varied in a wide range to {study the}
 effect on the nucleation rate. {I}t was found {that} the finite volume {effect is only noticeable for a } 
 narrow range of volumes and {that} it rapidly vanishes as the volume grows {so that CNT and dCNT are accurate above this threshold}. 

\begin{turnpage}
\begin{table*}
\caption{Properties of the capillary and extended cluster models as  function of the effective supersaturation ratio, $S_e$ with $V_T=4.5\times10^{6}\,\sigma^{-3}$. The absorbing wall was set to be $R_+=1.5\,R_*$. 
}\label{tab:j}
\begin{center}
\begin{tabular}{c|c|c|c|c|c|c|c}
  \hline\hline  
  && \multicolumn{6}{|c}{Modified Capillary Model} \\\hline
$S_e$ & ${j_{nc}}e^{-\Delta\beta F_*}$  & $R_{*E}$ & $\Delta N_*$ & $\Delta\beta F_*$ & 
  $\frac{j}{ {j_{nc}}}$& $\frac{j_{app}}{ {j_{nc}}}$ & $\rho_0$\\ \hline
 1.175 & 0.695 & 7.94 & 1.22$\times 10^3$   & 80.8 & 0.198 & 0.191 & 0.66503  \\
   1.5 & 0.770 & 3.3 & 84.4 		   & 14.3 & 0.623 & 0.772 & 0.66438  \\
   2   & 1.532 & 2.15 & 21.9 		   & 6.06 & 0.548 & 0.834 & 0.66412  \\
   2.5 & 2.758 & 1.84 & 12.8 		   & 4.44 & 0.393 & 0.672 & 0.66401  \\
\hline\hline
\end{tabular}
\vspace{0.5cm}

\begin{tabular}{c|c|c|c|c|c|c|c|c|c|c|c|c|c|c|c|c}
  \hline\hline  
  && \multicolumn{5}{|c|}{Extended model 0: $\rho_0=\rho_l$, $w=w_0$} & \multicolumn{5}{|c|}{Extended model 1: $\rho_0=\rho_{st^{(cap)}}$, $w=w_0$} & \multicolumn{5}{|c}{Extended model 2: $\rho_0=\rho_{st^{(plp)}}$, $w=w_{st^{(plp)}}$} \\ \hline
$S_e$ & $ {j_{nc}}e^{-\Delta\beta F_*}$  & $R_{*E}$ & $\Delta N_*$ & $\Delta\beta F_*$ & $\frac{j}{ {j_{nc}}}$& $\frac{j_{app}}{ {j_{nc}}}$
				   & $R_{*E}$ & $\Delta N_*$ & $\Delta\beta F_*$ & $\frac{j}{ {j_{nc}}}$& $\frac{j_{app}}{ {j_{nc}}}$
				   & $R_{*E}$ & $\Delta N_*$ & $\Delta\beta F_*$ & $\frac{j}{ {j_{nc}}}$& $\frac{j_{app}}{ {j_{nc}}}$\\
\hline
 1.175 & 0.695 & 8.03 & 933.25& 81.9922 &  0.0041  & 0.0028 & 8.02 &926.95 &81.5787 & 0.0063 & 0.0043 & 7.83 &787.32 &75.2840 & 2.9234 
  &1.7871\\
  1.5  & 0.770 & 3.39 & 51.68 & 14.9410 &  0.0313  & 0.0272 & 3.39 &51.37  &14.7634 & 0.0380 & 0.0332 & 3.09 &25.79  &10.6570 & 1.6432 & 1.2196\\
  2    & 1.532 & 2.16 & 10.57 & 5.8644  &  0.0903  & 0.1045 & 2.16 &10.52  &5.7515  & 0.1036 & 0.1205 & 1.82 &2.97   &2.8714  & 1.0116 &1.0912\\   
2.5    & 2.758 &1.73  & 4.62  & 3.4804  &  0.1605  & 0.2349 & 1.73 &4.60   &3.3993  & 0.1799 & 0.2648 & 1.39 &0.94   &1.2245  & 1.0186	  
  &1.0338\\
\hline\hline
 \end{tabular}
\vspace{0.5cm}

\begin{tabular}{c|c|c|c|c|c|c|c}
  \hline\hline  
 &&& \multicolumn{1}{|c|}{Extended model 0} & \multicolumn{2}{|c|}{Extended model 1} & \multicolumn{2}{|c}{Extended model 2} \\ \hline
$S_e$ &$\rho_l$ & $ {j_{nc}}/j_{\text{cnt}}$  &$\frac{j}{j_{\text{cnt}}}$
			      &$\rho_0$ & $\frac{j}{j_{\text{cnt}}}$
			      &$\rho_0$ & $\frac{j}{j_{\text{cnt}}}$\\
\hline
1.175 &0.665& 0.47   & 0.0235 & 0.6650 & 0.0358 & 0.6636 & 16.73 \\
1.5   &0.665& 1.15  & 0.3409 & 0.6644 & 0.4146 & 0.6635 & 17.91 \\
2     &0.665& 1.63  & 1.0449 & 0.6641 & 1.1988 & 0.6634 & 11.71 \\
2.5   &0.665& 2.57  & 2.3405 & 0.6640 & 2.6244& 0.6634 & 14.86 \\
\hline\hline
 \end{tabular}
\end{center}
\end{table*}
\end{turnpage}

\begin{acknowledgments}
The work of J.F.L is supported in part by the European Space Agency
under Contract No. ESA AO-2004-070 and by FNRS Belgium under Contract No.
C-Net NR/FVH 972. M.A.D. acknowledges support from the Spanish Ministry of Science and
Innovation (MICINN), FPI grant BES-2010-038422 (project AYA2009-10655).
\end{acknowledgments}

\bibliographystyle{unsrtnat}
\bibliography{bibliography/biblio}

\appendix
\section{Calculations for the piecewise-linear profile}
\subsection{The cumulative mass and metric}
Conforming to the postulated PLP, the cumulative mass can be computed for
$0\leq R\leq w$,
\begin{align}
 m(r)=&\Theta(R-r)\frac{\pi}{3w}(\rho_0- \rho_{ext}(R))r^3(4R-3r)\nonumber\\
 &+\Theta(r-R)\frac{\pi}{3w}(\rho_0-\rho_{ext}(R))R^4\nonumber\\
 &+V(r)\rho_{ext}(R)
 \label{eq:mass-Rltw-EM}
\end{align}
while for $R>w$ it becomes,
\begin{align}
 m(r)=&\Theta(R-w-r)V(r)(\rho_0-\rho_{ext}(R))\notag\\
 &+\Theta(r-(R-w))\Theta(R-r)\times\notag\\
 &\times\frac{\pi}{3w}(\rho_0-\rho_{ext}(R))
    \left(r^3(4R-3r)-(R-w)^4\right)\notag\\
 &+\Theta(r-R)\frac{\pi}{3w}(\rho_0-\rho_{ext}(R))\left(R^4-(R-w)^4\right)\notag\\
 &+V(r)\rho_{ext}(R)
 \label{eq:mass-Rgtw-EM}
\end{align}
with $V(r)=\frac{4\pi}{3}r^3$. According to these equations, the metric will present two
contributions for those clusters with $0\leq R\leq w$ and three addends
when $R>w$. In the first case,
\begin{align}
 g(R)=g_{i}(R)+g_{out}(R), \quad 0\leq R\leq w
 \label{eq:metric-Rltw-EM}
\end{align}
with,
\begin{align}
 g_{i}(R)=\int_0^{R}&\frac{1}{4\pi r^2\left(\rho_0-(\rho_0-\rho_{ext}(R))\frac{r-R+w}{w}\right)}\times\notag\\
 &\times\left(
    \begin{array}{l}
    \frac{4\pi}{3w}r^3(\rho_0-\rho_{ext}(R))\\
    \quad-\frac{\pi}{3w}(4R-3r)r^3\frac{\partial \rho_{ext}(R)}{\partial R}\\
    \qquad+V(r)\frac{\partial \rho_{ext}(R)}{\partial R}
    \end{array}
  \right)^2dr,\\
 g_{out}(R)=\int_R^{R_T}&\frac{1}{4\pi r^2 \rho_{ext}(R)}\times\notag\\
 &\times\left(
     \begin{array}{l}
     \frac{4\pi}{3w}R^3(\rho_0-\rho_{ext}(R))\\
     \quad-\frac{\pi}{3w}R^4\frac{\partial\rho_{ext}(R)}{\partial R}\\
     \qquad+V(r)\frac{\partial \rho_{ext}(R)}{\partial R}
     \end{array}
  \right)^2\,dr.
\end{align}
The first term concerns the cluster surface and the second one affects to the outside mass.
Although the exact solution exists and can be computed for these integrals, they are very crude.
Fortunately, we are interested in obtaining an analytical  approximation of the metric with
the aim of calculating a first order approximation of $Y$ for small clusters.
In this limit, it is a good estimation to consider
$\rho_{ext}(R)\sim\rho_{av}$ and $\partial\rho_{ext}(R)/\partial R\sim 0$, giving rise to
the expression already obtained for infinite systems,
\begin{align}
 g(R < w)\sim& \frac{8\pi}{15}\left(\frac{\rho_0-\rho_{av}}{w}\right)^2\frac{R^5}{\rho_{av}}
\label{eq:metric-approximation-EM}
\end{align}
The goodness of this approximation will be ultimately checked by comparison with the numerical results. For $R>w$ it also becomes an intractable equation
whose solution is extremely crude. Thus, writing it down would be meaningless since our interest on the metric expression relies on finding an approximation of this quantity for small clusters.

Once again, a canonical variable can be defined and it can be expanded about 
small clusters,
\begin{align}
 Y(R) \sim&\int_0^R \sqrt{g(R'<w)}dR'\nonumber\\
 \sim&\frac{2}{7}\left(\frac{\rho_0-\rho_{av}}{w}\right)\sqrt{\frac{8\pi}{15\rho_{av}}}R^{7/2}
\end{align}
where the approximation (\ref{eq:metric-approximation-EM}) has been used. 
Given that quantity, now we can get the approximation for the excess number of molecules
in the cluster, as we did previously. However, here $\Delta N$ has
to be evaluated carefully as there is no a simple relation with $R$ as in the 
MCP, but with the equimolar radius $R_E$,
\begin{align}
  \Delta N =& \int_0^{R_T} (\rho(\mathbf{r})-\rho_{ext}(R))d\mathbf{r}
  =\frac{4\pi}{3}R_E^3(\rho_0-\rho_{ext}(R)),
  \label{eq:delta-N-Re}
\end{align}
which in the case of the MCP is equivalent to $R$. After some manipulations one
arrives at,
\begin{align}
  R_E^3={\left(\frac{1}{4w}\left(R^4-\left(\max(R-w,0)\right)^4\right)\right)}.
  \label{eq:equimolar-R-EM}
\end{align}
Therefore, the excess number of molecules will satisfy the following approximation for
small clusters,
\begin{align}
\Delta N\sim&\frac{4\pi}{3}(\rho_0-\rho_{av})\frac{1}{4w}
  \left(\frac{7}{2}\left(\frac{w}{\rho_0-\rho_{av}}\right)
  \sqrt{\frac{15\rho_{av}}{8\pi}}\right)^{8/7}
\end{align}

\subsection{Free energy}
Now, as in the case
of the metric, the work of cluster formation will have two different expressions 
depending on whether $R>w$ or $0\leq R\leq w$. Fortunately, an exact equation can be
found in both cases, unlike for the metric. For $R>w$ equation (\ref{eq:W-EM-squaredGradient})
becomes,
\begin{align}
\Delta\beta F(R;w)=&\frac{4\pi}{3}(R-w)^3\beta\left( f(\rho_0)-f(\rho_{av})\right)\nonumber \\
&+ \left(1-\delta^3(R)\right)\,\beta\left( f(\rho_{ext}(R))-
f(\rho_{av})\right)\,V_T  \notag \\
&+4\pi\beta\left( \overline{\varphi}_0(R)w+ K\frac{%
(\rho_0-\rho_{ext}(R))^2}{2w}\right)R^2  \notag \\
&-4\pi\beta\left(2 \overline{\varphi}_1(R)w+ K\frac{%
(\rho_0-\rho_{ext}(R))^2}{2w}\right)Rw  \notag \\
&+4\pi\beta\left( \overline{\varphi}_2(R)w+ K\frac{(\rho_0-\rho_{ext}(R))^2}{6w}\right)w^2,
\label{eq:W-EM-Rgtw} 
\end{align}
with,
\begin{align}
\overline{\varphi}_k(R;w)=\frac{\int_{\rho_{ext}(R)}^{\rho_0}\left( f(x)-f(\rho_{av})\right)
   \left(x-\rho_{ext}(R)\right)^k dx}{(\rho_0-\rho_{ext}(R))^{k+1}}.
\label{eq:momentums-EM}   
\end{align}
It is easy to check that these expressions tend to their infinite-system counterparts
when the corresponding limit is taken into account. Besides, the model reproduces the same 
similarity when the small cluster limit is imposed, i.e. for $R<w$, 
\begin{align}
\beta  \Delta F(R;w)&=\left(1-\delta^3(R)\right)\,\beta
 (f(\rho_{ext}(R))-f(\rho_{av}))\,V_T \notag\\
&+\int_{0}^{R}4\pi\,r^2\times\nonumber\\
&\times \beta \Delta f\left(\rho_0-(\rho_0-\rho_{ext}(R))%
\frac{r-R+w}{w}\right)dr  \notag \\
&+\beta K\frac{2\pi}{3}R^3\left(\frac{\rho_0-\rho_{ext}(R)}{w}\right)^2,
\label{eq:F-EM-Rltw}
\end{align}
which can be eventually approximated by,
\begin{align}
\beta\Delta F(R;w)\sim& \beta K\frac{2\pi}{3}\left(\frac{%
\rho_0-\rho_{av}}{w}\right)^2R^3\nonumber\\
\sim& \frac{2\pi}{3}\beta K\left(\frac{\rho_0-\rho_{av}}{w}\right)^2\notag\\
&\times \left(\frac{7}{2}\left(\frac{w}{\rho_0-\rho_{av}}\right)\sqrt{\frac{15\rho_{av}}{8\pi}}\right)^{6/7}Y^{6/7},
\end{align}
equation which will be used to compute the nucleation rate subsequently, 
by using equation (\ref{eq:escape-rate-general}) with $\alpha=\frac{6}{7}$ and,
\begin{equation}
 \widetilde{F}_0=\frac{2\pi}{3}\beta K\left(\frac{\rho_0-\rho_{av}}{w}\right)^2\left(\frac{7}{2}\left(\frac{w}{\rho_0-\rho_{av}}\right)\sqrt{\frac{15\rho_{av}}{8\pi}}\right)^{6/7}.
\end{equation}

\end{document}